\documentclass[twocolumn]{aastex631}

\usepackage{amsmath}
\usepackage{multirow}
\usepackage{xcolor} 
\usepackage{hyperref}

\begin{document}

\title{Population demographics of white dwarf binaries with intermediate separations: Gaia constraints on post-AGB mass transfer}

\author[0000-0001-6970-1014]{Natsuko Yamaguchi}
\affiliation{Department of Astronomy, California Institute of Technology, 1200 E. California Blvd, Pasadena, CA, 91125, USA}

\author[0000-0002-6871-1752]{Kareem El-Badry}
\affiliation{Department of Astronomy, California Institute of Technology, 1200 E. California Blvd, Pasadena, CA, 91125, USA}

\author[0000-0001-9298-8068]{Sahar Shahaf}
\affiliation{Department of Particle Physics and Astrophysics, Weizmann Institute of Science, Rehovot 7610001, Israel}

\begin{abstract}
Astrometry from the Gaia mission has revealed a large population of white dwarf (WD) + main sequence (MS) binaries with periods of $100 - 1000\,$d. These systems have separations intermediate to predictions from standard binary evolution scenarios, challenging models of binary interaction and mass transfer. Because the selection function of Gaia astrometric catalogs is complex, the underlying population demographics of WD+MS binaries remain imperfectly understood. We present a forward-model of the AU-scale WD+MS binary population probed by Gaia that begins with a realistic binary population and incorporates a full model of Gaia mock observations and astrometric model fitting, as well as cuts employed in producing the Gaia astrometric catalog and selecting WD+MS binary candidates. We model the formation of AU-scale WD+MS binaries as the result of interaction when the WD progenitor is an AGB star. We test several models for the binaries' formation, including stable mass transfer with theoretically predicted stability criteria and two different formalisms for common envelope evolution. None of these models succeed in reproducing the observed component mass distributions or the absolute number of WD+MS binaries. The data are best reproduced by a model in which post-AGB binaries remain wide only if the accretor-to-donor mass ratio exceeds $\sim 0.4$. Our model allows us to constrain the intrinsic population demographics of intermediate-separation WD+MS binaries. The inferred period distribution is close to flat, with ${\rm d}N/{\rm d}P_{{\rm orb}}\propto P_{{\rm orb}}^{0.12}$, while the WD mass distribution is sharply peaked at $0.6\,M_{\odot}$. The model implies that $\sim 0.4\%$ of solar-type stars have WD companions with periods of $100 - 1000\,$d. 
\end{abstract}


\section{Introduction} \label{sec:intro}

\defcitealias{El-Badry2024OJAp_b}{E24}

Close white dwarf (WD) + main sequence (MS) binaries are end products of binary interactions that occurred when the WD progenitor was a giant. They are therefore useful probes of the physics of mass transfer (MT) processes which play an important role in the evolution of many types of binaries.

The third data release (DR3) of the Gaia mission introduced the non-single stars (NSS) catalog, which included orbital solutions for an unprecedented number of binaries. In particular, it contains astrometric orbital solutions for over 160,000 sources, leading to the discovery of new populations of luminous stars orbited by black holes, neutron stars, and WDs \citep{El-Badry2023MNRAS, El-Badry2023MNRASb, Chakrabarti2023AJ, El-Badry2024OJAp, Shahaf2023MNRAS, Shahaf2024MNRAS}.  
To identify astrometric binaries likely to contain compact objects, \citet{Shahaf2019MNRAS} developed a ``triage" scheme. It rests on a quantity called the astrometric mass ratio function (AMRF), which depends only on observables from Gaia and is larger for systems hosting dark companions than for systems hosting luminous companions. Using this technique, \citet{Shahaf2024MNRAS} identified a sample of over 3000 high-probability WD+MS binaries with orbital periods ranging from $\sim 100-1000\,$days, corresponding to AU-scale separations. While post-interaction binaries with comparable separations were known before Gaia \citep[e.g.][]{Kawahara2018AJ, Masuda2019ApJL, Oomen2018A&A, Jorissen2019A&A, Escorza2019A&A}, the scale of the population was not appreciated because most photometric and spectroscopic searches are more sensitive to short than to long orbital periods. 

The abundance of these AU-scale WD+MS binaries was unexpected, because standard binary evolution models predicted few such systems to exist. While the binaries' separations are too close to have avoided interaction when the WD progenitor was on the asymptotic giant branch (AGB) -- implying they are post-MT systems -- their orbits are significantly longer than predicted by models of dynamically unstable MT (i.e. common envelope evolution, CEE; e.g. \citealt{Paczynski1976IAUS, Ivanova2013A&ARv}). An alternative possibility is that the systems are products of stable MT. While classical models predicted MT  from convective donors more massive than their companions to become unstable \citep[e.g.][]{Paczynski1965AcA, Paczynski1969ASSL, Hjellming1987ApJ}, recent work has shown that MT may be stable for a wider range of progenitor mass ratios \citep[e.g.][]{Ge2020ApJ, Temmink2023AA}. Although several models predict MT from AGB donors to be stable for a wide range of mass ratios, they generically predict extremely high mass transfer rates \citep[$\gtrsim 10^{-2}\,M_{\odot}\rm yr^{-1}$; e.g.][]{Temmink2023AA, Yamaguchi2024PASP}. It is not clear that binaries can survive such high mass transfer rates, since they may lead to the inflation of the accretor and ultimately, CEE \citep[e.g.][]{Lau2024}.

Besides the objects discovered by Gaia, there are several other related classes of post-interaction binaries with comparable orbital periods, such as WD symbiotic binaries  \citep[e.g.][]{Lu2006MNRAS}, post-AGB binaries \citep[e.g.][]{VanWinckel2025}, and barium stars \citep[e.g.][]{Escorza2019A&A}, whose orbits are difficult to explain with existing binary interaction models. Insights into the mass transfer process from the Gaia WD+MS sample can also shed light on the formation history of these objects.

While several works have attempted to model the evolution of wide post-interaction WD+MS binaries \citep[e.g.][]{Belloni2024A&A, Yamaguchi2024MNRAS}, a population-level study requires an understanding of the selection function. This is made challenging by the fact that Gaia DR3 does not contain epoch astrometry or orbital solutions for all sources. Instead, each source goes through a multi-step pipeline, where cuts are made to remove potentially spurious solutions and where a full astrometric binary model is only attempted if simpler models do not provide a sufficiently good fit to the data. This means that while the published orbits are largely reliable \citep[e.g.][]{Yamaguchi2024PASP}, many true binaries do not make it into the NSS catalog. Inference of population-level properties of the WD+MS binary sample, such as period, eccentricity, and mass ratio distributions, has thus far remained limited by the poorly-understood selection function of the NSS catalog.

Recently, \citet{El-Badry2024OJAp_b} (\citetalias{El-Badry2024OJAp_b} henceforth) used a forward-model of Gaia epoch astrometry and astrometric solution processing to approximate the Gaia astrometric orbit selection function. Starting with a model of the Galactic binary star population, they produced synthetic Gaia epoch astrometry for each source. They then ran the synthetic astrometry through the same cascade of astrometric models used in producing the NSS catalog. They showed that the distributions of orbital and stellar parameters of the binaries in the resulting mock catalog are in good agreement with those of the real catalog, implying that their selection function is a good approximation of reality.

In this work, we extend the work of \citetalias{El-Badry2024OJAp_b}, focusing on the WD+MS binary population. We self-consistently forward-model the creation of the Gaia DR3 astrometric orbit sample, the selection of WD+MS binary candidates via the AMRF, and potential contamination of the sample from hierarchical triples, tuning the assumed underlying WD+MS binary population to match the final observed sample from \citet{Shahaf2024MNRAS}. In doing so, we constrain the occurrence rate and population demographics of WD+MS binaries.

The rest of this paper is organized as follows: In Section \ref{sec:forward_model}, we summarize the steps taken by \citetalias{El-Badry2024OJAp_b} to create a mock Gaia NSS catalog and describe our addition of hierarchical triples to their model. In Section \ref{sec:amrf_sample}, we provide an overview of the triage technique employed by \citet{Shahaf2024MNRAS} to construct the observed sample of WD+MS binaries using orbital solutions from Gaia DR3. We also describe its application to our simulated binaries. In Section \ref{sec:e24_model}, we compare properties of the resulting mock sample of WD+MS binaries obtained using the original model from \citetalias{El-Badry2024OJAp_b} to those of the observed sample. Based on this, in Section \ref{sec:modifications}, we describe several modifications to the \citetalias{El-Badry2024OJAp_b} model and different models of the orbital evolution under MT to attempt to better match the data. In Section \ref{sec:results_modified}, we compare the models' predictions to data and discuss completeness fractions and the underlying parameter distributions of WD+MS binaries. Finally, we conclude in Section \ref{sec:conclusion}. 

\section{Forward model of the Gaia DR3 NSS catalog} \label{sec:forward_model}

We begin by summarizing the \citetalias{El-Badry2024OJAp_b} model of Gaia astrometric orbit catalogs. We refer readers to \citetalias{El-Badry2024OJAp_b} for a more detailed description. 

\subsection{Initial binary population} \label{ssec:initial_binary_pop}

Stellar ages, metallicities, positions, and kinematics are generated using \texttt{Galaxia} \citep{Sharma2011ApJ}. As 99$\%$ of the systems in the NSS catalog are found within 2 kpc of the Sun, only sources within this distance are modeled.

Since \texttt{Galaxia} does not include binaries, our model uses \texttt{COSMIC} \citep{Breivik2020ApJ} to generate a zero-age binary population with parameter distributions from \citet{Moe2017ApJS}. Binaries are matched to stars in the \texttt{Galaxia} population by their primary masses and placed at the same position. Binary orientations and phases are assigned randomly. G-band magnitudes are calculated from \texttt{MIST} isochrones \citep{Morton2015ascl, Choi2016ApJ} for each binary component. We calculate extinctions using the \texttt{combined19} dust map in the \texttt{mwdust} package \citep{Bovy2016ApJ}, which combines dust maps from \citet{Drimmel2003A&A, Marshall2006A&A, Green2019ApJ}. 

\subsection{Formation of compact objects} \label{ssec:CO_formation}

Binaries with primaries with initial masses above $8\,M_{\odot}$, which have terminated their evolution as black holes or neutron stars, are removed under the assumption that most of them are broken up or shrink to very tight orbits via CEE. Stars with initial masses below $8\,M_{\odot}$ are assumed to end their lives as WDs. \citetalias{El-Badry2024OJAp_b} predicted WD masses using the empirical initial-final-mass relation (IFMR) of \citet{Weidemann2000AA}, which is calibrated to observations of single WDs. By default, we continue to use this relation in this work, but have verified that using a more updated IFMR from \citet{Cunningham2024MNRAS} has a minimal effect on our final results.

\subsection{Post-interaction separations} \label{ssec:post_interaction_systems}

\citetalias{El-Badry2024OJAp_b} assumed that binaries with initial separations $< 2\,$AU experience mass transfer which efficiently shrinks and circularizes their orbits such that they end up in 1-day orbits with zero eccentricities which are not detectable with astrometry. Meanwhile, they assumed that 10\% of orbits with initial separations between $2$ and $6\,$AU end up in wider orbits with final separations half of their initial separations and moderate final eccentricities randomly selected between 0 and 0.2 (the remaining 90\% end up in 1-day circular orbits). The $2\,$AU boundary was chosen to roughly distinguish between systems that interact with donors on the red giant branch (RGB) versus the AGB. This was motivated by recent studies of the Gaia WD+MS binary sample which have found that it is possible to from relatively wide post-common envelope binaries (PCEBs) if mass transfer begins during a thermal pulse of an AGB donor \citep[e.g.][]{Belloni2024A&A, Yamaguchi2024PASP}. On the other hand, there are also recent studies showing that MT from AGB donors may remain stable for relatively unequal mass ratios, which could lead to wider post-MT orbits than expected for classical PCEBs \citep[e.g.][]{Ge2020ApJ, Temmink2023AA}. These assumptions regarding the orbital evolution due to MT are modified in this work (Section \ref{ssec:fshrink_models}). 

\citetalias{El-Badry2024OJAp_b} did not model the formation of low-mass helium WDs that are formed through stable mass transfer from donors on the first giant branch \citep[e.g.][]{Rappaport1995MNRAS, Marsh1995MNRAS, Brown2016ApJ, Li2019ApJ, El-Badry2022MNRAS, Garbutt2024MNRAS}. Although low-mass WDs formed via stable mass transfer are common in samples of WD binaries selected via UV excess, they are almost entirely absent in the AMRF sample, which is only sensitive to systems with $M_{\rm WD}/M_\star \gtrsim 0.6$ \citep{El-Badry2024NewAR}. 

\subsection{Mock observations}

We produce mock observations of each source using the \texttt{gaiamock}\footnote{https://github.com/kareemelbadry/gaiamock} package, which is described in detail by \citetalias{El-Badry2024OJAp_b}. In brief, the code produces simulated epoch astrometry following the Gaia scanning law, models the effect of binarity on the epoch astrometry following \citet{Lindegren2022}, and implements an empirical noise model based on the residuals of DR3 epoch astrometry for well-behaved sources from \citet{Holl2023A&A}. The simulated epoch astrometry is then fit with the same cascade of astrometric models used in Gaia DR3 \citep{Halbwachs2023A&A}, ultimately producing a mock NSS catalog that can be compared to the observed NSS catalog.

\begin{figure*}
    \centering
    \includegraphics[width=0.95\linewidth]{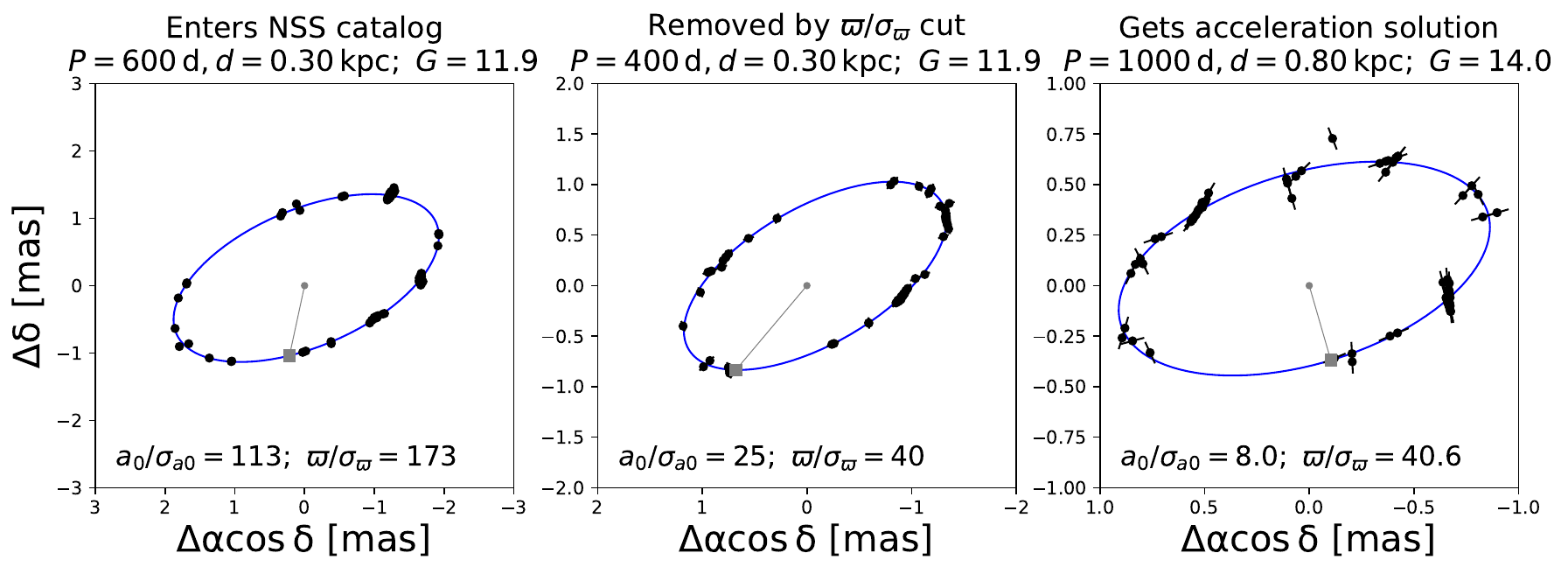}
    \caption{Examples of orbits and their mock observations. In all cases, we consider a $1.0\,M_{\odot}$ luminous primary and a dark $0.6\,M_{\odot}$ secondary in a slightly eccentric ($e=0.1$) orbit. Different orbital periods and distances are considered in each panel. The ratio of semi-major axis and parallax to their respective errors are quoted on the bottom of each panel; quality cuts on these parameters determine whether each binary receives an orbital solution. The leftmost panel is most representative of a typical system in the AMRF sample and receives an orbital solution. The middle one is excluded from the NSS catalog due to having $\varpi/\sigma_{\varpi}$ that is too low, despite the orbit being fairly well constrained. The rightmost orbit gets an acceleration solution and is therefore not fit with an orbital solution.}
    \label{fig:example_orbits}
\end{figure*}

In Figure \ref{fig:example_orbits}, we show several simulated WD+MS binaries, along with their mock observations. The leftmost panel shows a typical WD+MS binary in the AMRF sample with orbital period, $P_{\rm orb}$, of $600\,$d at a distance of $300\,$pc. This system enters the mock NSS catalog with a full astrometric orbital solution. In the middle panel is a case of an orbit that receives an orbital solution but is excluded from the final catalog because it does not pass a quality cut imposed on the parallax signal-to-noise ratio to remove spurious solutions ($\varpi/\sigma_{\varpi} > 20000/P_{\rm orb}$; \citealt{Halbwachs2023A&A}). Finally, the orbit shown in the right panel, placed at a larger distance, only receives a 7-parameter acceleration solution and therefore also does not enter the NSS catalog.

\subsection{Triples} \label{ssec:triples}

\begin{figure}
    \centering
    \includegraphics[width=0.95\linewidth]{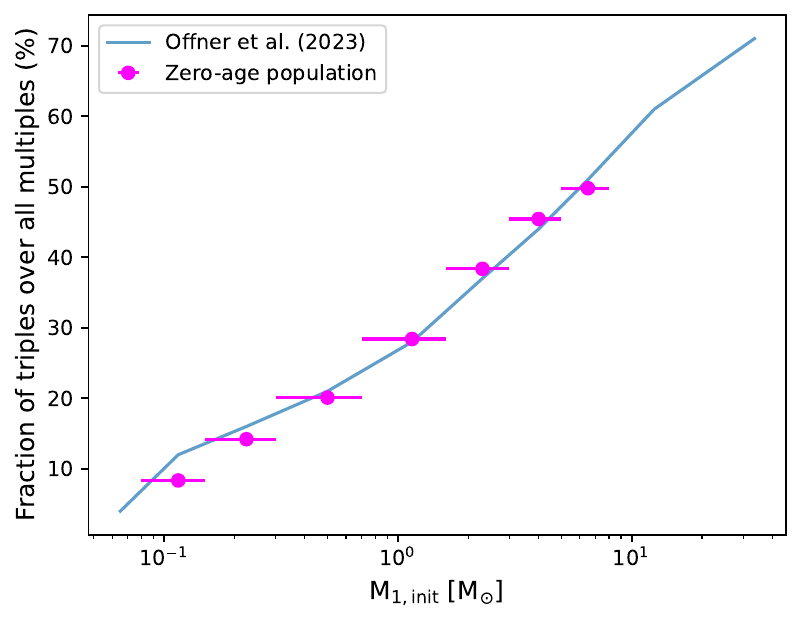}
    \caption{Fraction of the simulated zero-age binary population that are in hierarchical triples as a function of the true primary mass (i.e. maximum mass of the two/three components in the system). In blue is the inputted triple fraction from \citet{Offner2023ASPC}.}
    \label{fig:triple_frac}
\end{figure}

Because an inner binary of two MS stars has a higher mass-to-light ratio than a single star, hierarchical triples are the most important possible contaminant for an AMRF-selected WD sample (Section \ref{sec:amrf_sample}). To study their contribution to the final sample, we modify the \citetalias{El-Badry2024OJAp_b} model to include triples. We assign an outer tertiary to a subset of the zero age binary population with probability that depends on the primary mass. The triple fraction as a function of primary mass is obtained from Table 1 of \citet{Offner2023ASPC}. Since we are assigning triples to existing binaries, we use the fraction of triples and higher-order multiples in all multiple star systems (in the terminology adopted by \citealt{Offner2023ASPC}, this is THF/MF). For simplicity, in this step, we define the primary as the most massive component of the inner binary. While this is not exact (as the outer tertiary may end up most massive), Figure \ref{fig:triple_frac} shows that this procedure results in a triple fraction close to the observational constraints of \citet{Offner2023ASPC}.

For systems selected to be in triples, we draw masses of the outer tertiary assuming a uniform distribution of the mass ratio of the outer tertiary over the inner binary. The eccentricities of the outer orbits are sampled from a thermal distribution, and the periods from a log-normal distribution with a mean in log($P/\mathrm{d}$) of 4.54 and standard deviation of 2.4 \citep{Tokovinin2014AJ}. This process is repeated until each system meets the stability criteria for hierarchical triples from \citet{Naoz2013MNRAS} and \citet{Mardling2001MNRAS} (as summarized in Section 2.3 of \citealt{Shariat2023ApJL}). 

For the astrometric processing, we assume that if the angular size of the outer orbit (= semi-major axis/distance) is more than 1'' on the sky, the outer tertiary is resolved and will receive its own 5-parameter solution \citep{Nagarajan2024PASP}. Therefore, the inner binary of such a system is processed independently. If the size of the outer orbit is smaller than 1'' and its orbital period is less than 1000 d, the system is modeled as a binary made up of the inner binary and outer tertiary. The total mass and luminosity of the inner binary is taken as the mass and luminosity of one component. Such a system may receive an astrometric binary solution that can contaminate the AMRF sample. We discard triples with outer orbital periods longer than 1000 d and luminous tertiaries: these periods are too long for the outer orbit to be constrained in Gaia DR3, and the presence of the tertiary will disturb the astrometric measurements of the inner orbit \citep[e.g.][]{Lindegren2022}. We also neglect systems where both components of the inner binaries have evolved into WDs, since our modeling is unlikely to reliably track the mass transfer that preceeds the second WD's formation.

As we discuss in Section \ref{sec:results_modified}, the color excess cut ultimately removes the vast majority of triples, meaning that the final WD+MS binary sample is relatively insensitive to our assumptions about the triple population. However, modeling of triples is important for interpretation of the AMRF sample, because some genuine WD+MS binaries are excluded by cuts designed to filter out triples. 

For most of the subsequent analysis, we only consider triples and WD+MS binaries, excluding all other binaries which host two luminous companions. The latter are almost entirely removed by cuts on the AMRF and thus do not enter the final simulated WD+MS binary sample.

\begin{figure}
    \centering
    \includegraphics[width=0.97\linewidth]{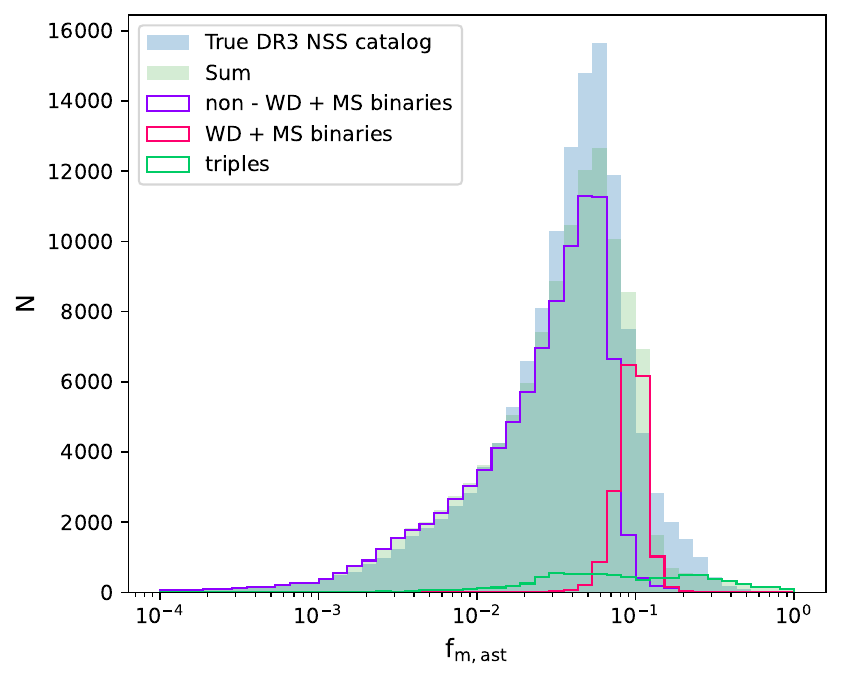}
    \caption{Distributions of the astrometric function, $f_{\rm m,ast}$, for the orbital solutions in the simulated empirical (green) and true (blue) NSS catalogs. Contributions from the different components in the simulated catalog are shown with unfilled histograms. We see that the total number of orbital solutions as well as the $f_{\rm m,ast}$ distribution of the mock and true catalog are in rough agreement with each other.}
    \label{fig:fm_mast_no_giants}
\end{figure}

In Figure \ref{fig:fm_mast_no_giants}, we show the distributions of astrometric mass function, $f_{\rm m,ast}$, of the orbital solutions in the mock NSS catalog resulting from our empirical model (described in Section \ref{sssec:empirical_model}) as well as the true NSS catalog published in Gaia DR3. To isolate MS primaries and exclude giants, we only plot sources in the observed catalog which are also present in the \texttt{binary\_masses} catalog and only consider simulated binaries with primary radii $< 2\,R_{\odot}$. $f_{\rm m,ast}$ is given by
\begin{equation}
    f_{\rm m,ast} = \left(\frac{\alpha}{\varpi}\right)^3 \left( \frac{P_{\rm orb}}{1\mathrm{yr}} \right)^{-2}.
\end{equation}
This quantity is useful for identifying both spurious astrometric solutions \citep[][]{Halbwachs2023A&A} and systems with compact object companions. 

In the simulated sample, most of the low-$f_{\rm m,ast}$ solutions are non-WD+MS binaries, while the highest-$f_{\rm m,ast}$ solutions are triples. WD+MS binaries lie in between these populations, with a median  $f_{\rm m,ast}$ of $0.09$. In total (not excluding giants), there are 156,441 orbital solutions in the mock catalog, which is $7\%$ fewer than the 168,000 in the observed catalog. For comparison, the original \citetalias{El-Badry2024OJAp_b} 
catalog contained $18\%$ fewer sources than the true catalog. The difference is primarily due to the larger number of WD+MS binaries in our model. As we will show (Section \ref{sec:results_modified}), the typical WD mass in the sample is lower than assumed by \citetalias{El-Badry2024OJAp_b}. As a result, a smaller fraction of WD+MS binaries enter the AMRF sample than assumed in their work, and so a larger underlying WD+MS binary population is required to explain the same AMRF sample. The observed and simulated  $f_{\rm m,ast}$ distributions are in reasonably good agreement. WD+MS binaries make up $14\%$ of sources in the mock catalog, and $15\%$ of these make it into the final WD+MS sample. This implies that the observed NSS catalog likely contains many WD+MS binaries that have yet to be identified. 
 
\section{AMRF sample of WD+MS binaries} \label{sec:amrf_sample}

We apply the same cuts to mock binary catalogs as were applied to the NSS catalog by \citet{Shahaf2024MNRAS}. We first calculate the AMRF, $\mathcal{A}$: 
\begin{equation}  \label{eqn:AMRF_1}
    \mathcal{A} = \frac{\alpha}{\varpi}\left( \frac{M_1}{M_{\odot}} \right)^{-1/3}\left( \frac{P_{\rm orb}}{\mathrm{yr}} \right)^{-2/3}
\end{equation}
where $\alpha$ is the angular photocentric semi-major axis, $\varpi$ is the parallax, $M_1$ is the primary mass, and $P_{\rm orb}$ is the orbital period.
The AMFRF can be rewritten in terms of the mass ratio, $q$, and G-band flux ratio, $\mathcal{S}$, of the system: 
\begin{equation} \label{eqn:AMRF_2}
    \mathcal{A} = \frac{q}{(1+q)^{2/3}} \left( 1 - \frac{\mathcal{S}(1+q)}{q(1+\mathcal{S})} \right)
\end{equation} 
where the ratios are defined as the photometric secondary over primary. 

Gaia provides constraints on all parameters in equation \ref{eqn:AMRF_1}. \citet{Shahaf2024MNRAS} obtained primary mass estimates from the \texttt{binary\_masses} catalog \citep{GaiaCollaboration2023A&A}, while they calculated $\alpha$ from the Thiele-Innes coefficients reported in the \texttt{nss\_two\_body\_orbit} catalog. 

For our simulated systems, we use the true primary masses from our model, implicitly assuming that the masses reported in the \texttt{binary\_masses} catalog are accurate and do not suffer systematic biases. We assume a fixed uncertainty of $0.06\,M_{\odot}$ in $M_1$, which is the average uncertainty in the observed sample. We remove simulated binaries with primary radii $> 2\,R_{\odot}$, as evolved stars are not included in the \texttt{binary\_masses} catalog and are thus also excluded by \citet{Shahaf2024MNRAS}. 

Using theoretical mass-luminosity relations for a range of primary masses, \citet{Shahaf2024MNRAS} calculate boundaries in the $\mathcal{A}-M_1$ space to define several classes corresponding to the most likely types of secondaries: (1) a single MS star, (2) an inner binary of two MS stars (i.e. hierarchical triple), and (3) a dark companion. As a starting point, they isolated systems in the latter two classes (i.e. inconsistent with hosting a single MS companion), which they call the ``non-\textit{class I}" sample. For details regarding the application of this technique to the orbits from the real NSS catalog, we refer readers to \citet{Shahaf2023MNRAS, Shahaf2024MNRAS}. 

\subsection{Color excess cut} \label{ssec:color_excess_cut}

\begin{figure}
    \centering
    \includegraphics[width=0.99\linewidth]{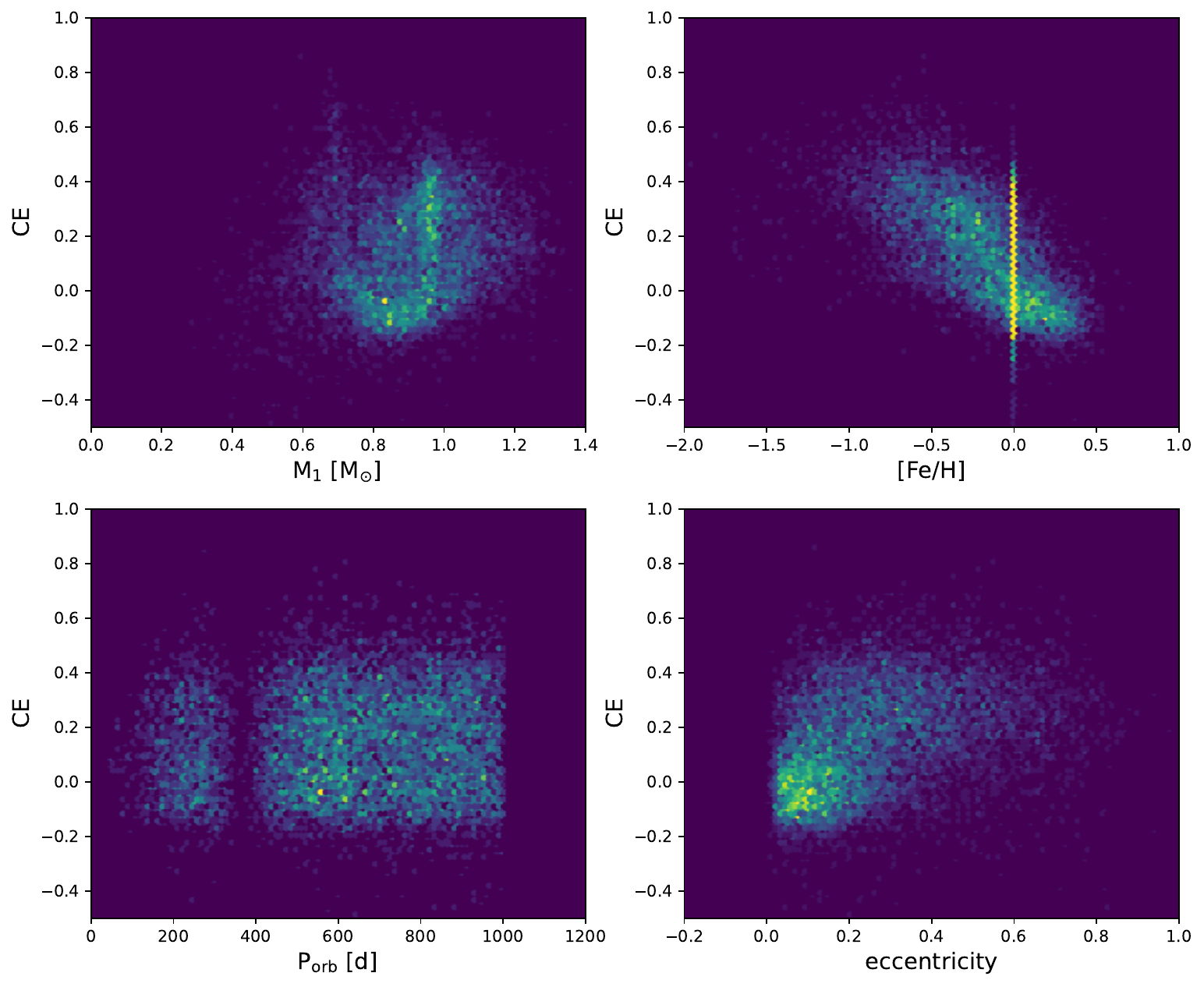}
    \caption{The four panels show 2D binned plots of the color excess against various stellar and orbital parameters of all sources in the non-\textit{class I} catalog of \citet{Shahaf2024MNRAS}.}
    \label{fig:real_CE_v_params}
\end{figure}

To reduce contamination from hierarchical triples, \citet{Shahaf2024MNRAS} removed systems from the non-\textit{class I} sample that are redder than expected for a single luminous companion. They did this by calculating the ``color excess'' (CE) of each source:
\begin{equation}
    \Delta(B-I) = (B-I)_{\rm observed} - (B-I)_{\rm expected}
\end{equation}
where $(B-I)_{\rm observed}$ and  $(B-I)_{\rm expected}$ are the de-reddened observed and theoretically expected $B-I$ colors. 

\citet{Shahaf2024MNRAS} used observed B, V, and I band magnitudes generated using the BP/RP spectra from the Gaia \texttt{synthetic\_photometry\_gspc} catalog and used 3D dust extinction maps \citep{Green2019ApJ, Lallement2019A&A} to correct for reddening. To calculate $(B-I)_{\rm expected}$, they interpolated PARSEC isochrones with a fixed age of 2 Gyrs on metallicity and the absolute V band magnitude using the \texttt{stam}\footnote{github.com/naamach/stam} package \citep{Hallakoun2021MNRAS}. They assumed a fixed age of 2 Gyr because the ages of the observed systems are unknown. In Section \ref{sssec:color_excess_dist}, we discuss the effect of this assumption on the simulated sample. They obtained metallicity estimates from \citet{Zhang2023MNRAS} who created an empirical forward model of the Gaia XP spectra to obtain stellar atmospheric parameters for 220 million sources. For the $\sim 20\%$ of sources flagged by \citet{Zhang2023MNRAS} to have unreliable stellar parameters, \citet{Shahaf2024MNRAS} set [Fe/H] = 0.0, with an uncertainty of 0.25 dex. 

For each source, \citet{Shahaf2024MNRAS} calculated $10^4$ samples of the color excess, each time drawing observables from uncorrelated Gaussian distributions with standard deviations set to their errors. Sources where the fraction of realizations with $\Delta(B-I) > 0$ exceeded 64\% (corresponding to the value above which the infrared excess is greater than than 1\%) were removed\footnote{An error in the code used by \citet{Shahaf2024MNRAS} led this threshold to be 56\% in the first-published catalog. This has since been corrected, and we compare our models to this updated catalog \citep{Shahaf2025MNRAS}.}, leaving behind $\sim 3000$ high probability WD+MS binaries in the final ``no-color excess'' (NCE) sample. 

For the simulated population, we predict the B, V, and I magnitudes of each system by interpolating on PARSEC isochrones and assuming the sources' true age and metallicity. Our grid of isochrones has log(age) ranging from 6.5 to 10.5 in steps of 0.1 dex and [M/H] from -2.0 to 1.0 in steps of 0.1 dex. We calculate the observed apparent magnitudes from the total flux of all components, which are then extincted/reddened according to the true distance to the source. We use the \texttt{dustmaps} \citep{2018JOSS....3..695M, Green2019ApJ} and \texttt{mwdust} python package (\citealt{Bovy2016ApJ, Drimmel2003A&A, Marshall2006A&A, Green2019ApJ}) to predict extinctions. We calculate the uncertainties in these apparent magnitudes via interpolation of an exponential fit to the flux errors as a function of the V band fluxes of sources in the observed non-\textit{class I} sample. We also fit residuals to the best-fit curve with an exponential, which we used to add noise to the interpolated errors, drawn from a normal distribution with a standard deviation given by the residual model. Finally, we convert the apparent magnitudes to ``observed" absolute magnitudes using the measured parallaxes from the mock astrometric fits.

We calculate $(B-I)_{\rm expected}$ using the same code as \citet{Shahaf2024MNRAS}, similarly assuming a fixed age of 2 Gyr. For the majority of sources, we use the true values of [Fe/H] generated by \texttt{Galaxia} and assume a fixed uncertainty of 0.04 dex, the median value for the observed sample. To match the observed sample, we randomly select $20\%$ of sources to have unreliable metallcities; for these, we set [Fe/H] = 0 and an elevated uncertainty of 0.25 dex. 

When calculating these absolute magnitudes and their uncertainties, \citet{Shahaf2024MNRAS} incorrectly used parallaxes and parallax uncertainties from the \texttt{gaia\_source} catalog, as opposed to those from the NSS catalog, which accounts for binary motion. While the parallax distribution is minimally affected by this, the reported uncertainties are, on average, a factor of 2.5 larger in the \texttt{gaia\_source} catalog. To remain consistent with their method, we therefore increase our mock parallax uncertainty measurements by the same factor in this calculation. 

\subsubsection{Color excess distribution} \label{sssec:color_excess_dist}

\begin{figure*}
    \centering
    \includegraphics[width=0.8\linewidth]{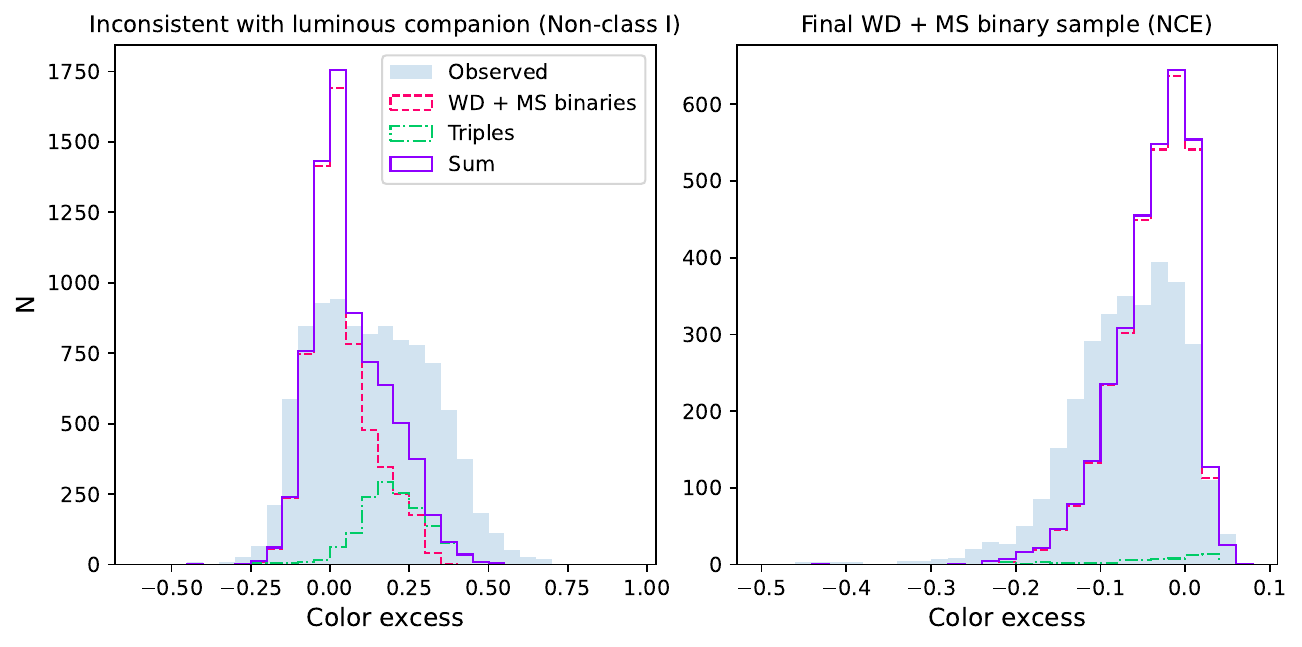}
    \caption{Distribution of color excess for systems in the non-\textit{class I} sample (\textit{Left}) and those that end up in the final NCE sample (\textit{Right}). On average, triples have larger color excess and are therefore effectively excluded from the final sample. However, we also find that more than half of true WD+MS binaries are removed in the process. The distribution for the observed non-\textit{class I} sample is significantly broader than that of our model, possibly due to inaccuracy in the measured metallicities (Section \ref{sssec:color_excess_dist} and Appendix \ref{ssec:appendix_CE_met_spread}).}
    \label{fig:color_excess_dist}
\end{figure*}

Figure \ref{fig:real_CE_v_params} shows the reported color excess of the observed non-\textit{class I} sample as a function of $M_1$, [Fe/H], $P_{\rm orb}$, and eccentricity. Recall that \citet{Shahaf2024MNRAS} set [Fe/H] = 0.0 for $\sim 20\%$ of sources with unreliable metallicites, causing a vertical streak at [Fe/H] = 0.0. Several other features stand out. Firstly, we observe a significant anti-correlation between color excess and [Fe/H]: essentially all sources with [Fe/H] $< -0.5$ have significant color excess. The source of this trend is uncertain, with one possibility being that the assumed relationship between color and metallicity is too strong in the PARSEC models (i.e. metal-poor stars are predicted to be bluer than they really are). Metal-poor stars are old, and thus on average are more evolved than they would be at the assumed age of 2 Gyr. This means that their color excess will be overestimated on average, and that the observed sample is likely to be biased against metal-poor WD+MS binaries.

Second, there is a correlation between color excess and eccentricity. Hierarchical triples are expected to have higher eccentricities on average than WD+MS binaries, so this trend likely reflects an increasing fraction of triples at high color excess. Further exploration of the trend can be found in Appendix \ref{ssec:appendix_triplefrac_met}. 

We also observe a cloud of sources with primary masses of $0.6 - 1.2\,M_{\odot}$ with systematically negative color excess, suggesting that the PARSEC models predict stars to be redder than they are in reality.

Since we use PARSEC isochrones to calculate both the true and expected colors of the simulated binaries, the color excess distribution of our simulated WD+MS binaries peaks at zero, as expected. To fully reproduce the observed relation between color excess and primary mass would require identifying and correcting the underlying cause of the inaccurate colors predicted by the models, which we do not attempt to do in this work. Instead, we apply an empirical correction to the color excess of our simulated systems. This is detailed in Appendix \ref{ssec:appendix_CE_correction}. We note that the distributions of key parameters in the observed and modeled populations are comparable prior to the cut on color excess (i.e. the non-\textit{class I} sample), so details of this correction do not play a critical role in our conclusions. However, the correction does affect the total number of systems that enter the final sample and therefore adds uncertainty to the inferred occurrence rate of wide post-MT systems. 

The distributions of color excess for systems that enter the simulated (Section \ref{sssec:empirical_model}) and observed non-\textit{class I} and final NCE samples are plotted in Figure \ref{fig:color_excess_dist}. On average, triples have significant non-zero color excess, which results in the vast majority of them being excluded from the final sample. At the same time, around half of the simulated WD+MS binaries also have non-zero color excess and are excluded, which is primarily due to the fixed age assumption made in the calculation of color excess (Section \ref{ssec:color_excess_cut}) and suggests a less stringent cut may have been possible. In addition, the observed distribution is broader. 

A possible explanation is that there is a discrepancy between the true and measured metallicities. This may also be responsible for the difference in the metallicity distributions of the observed and model samples (see discussion around Figure \ref{fig:nce_dist}). We note that the \citet{Zhang2023MNRAS} metallicities were calculated assuming \texttt{gaia\_source} parallaxes (as opposed to those from the NSS catalog) and did not account for the presence of light from luminous companions in the XP spectra. Since the uncertainties on these parallaxes are underestimated \citep{El-Badry2025arXiv}, it is likely that the metallicities are less accurate than their uncertainties suggest. The effect of introducing such a discrepancy is explored in Appendix \ref{ssec:appendix_CE_met_spread};  it does not significantly affect the distributions of the other parameters in the final NCE sample. However, a broader color excess distribution reduces the total number of systems that enter the final sample, adding another source of uncertainty on the inferred fraction of post-AGB MT systems that end up in wide orbits.

\subsubsection{AMRF distributions}

\begin{figure*}
    \centering
    \includegraphics[width=0.9\linewidth]{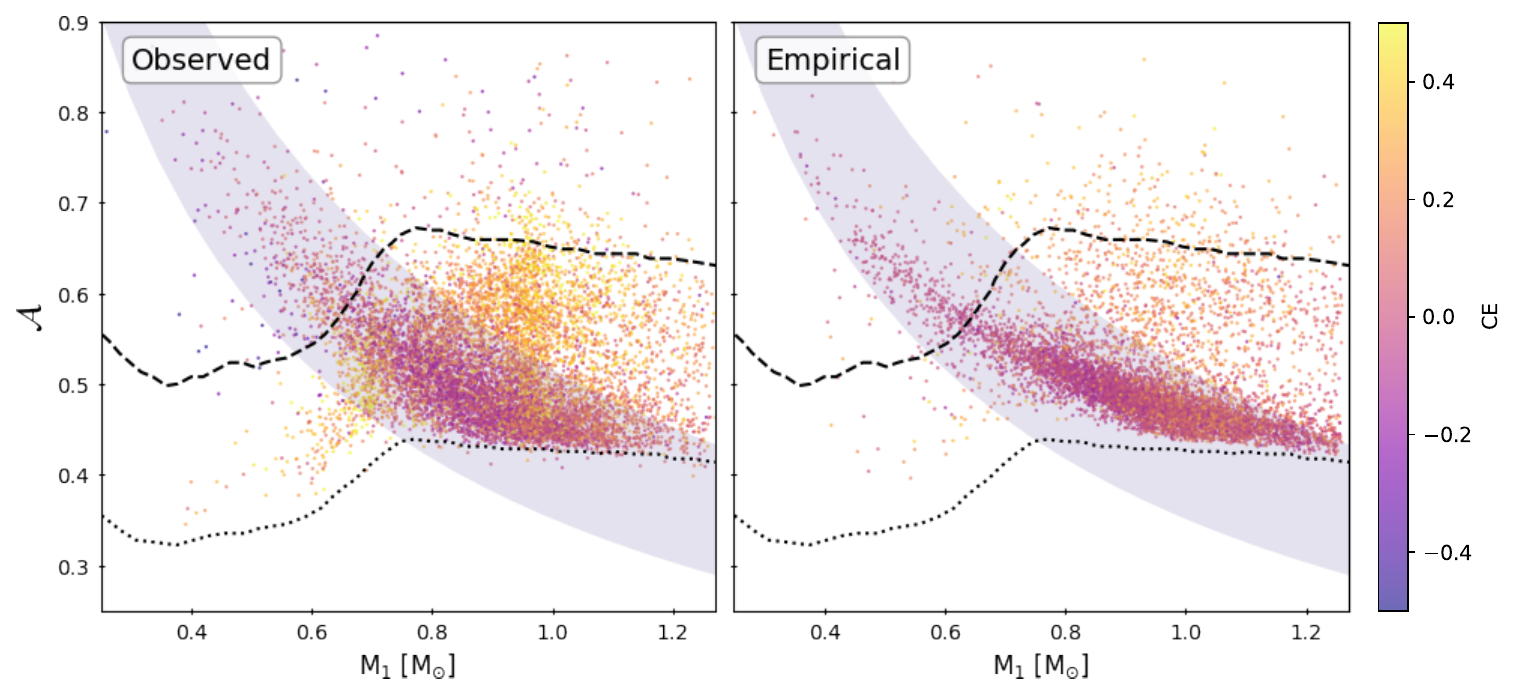}
    \caption{AMRF against primary mass for the observed (left) and mock (empirical model; right) non-\textit{class I} samples. The points are colored by their color excess. The shaded region is the expected location for systems hosting dark secondaries with masses between $0.45-0.75\,M_{\odot}$ (i.e. WD+MS binaries). The dashed and dotted lines are upper limits to systems consistent with hosting a single MS secondary and an inner binary of two MS stars, respectively. In both samples, there is a group of sources with near-zero color excess located in the shaded region, corresponding to WD+MS binaries. Sources with large positive color excess values are most likely triples, which are rarer in our simulated population than in the observed sample.}
    \label{fig:amrf_m1}
\end{figure*}

In Figure \ref{fig:amrf_m1}, we show the AMRF and primary mass for sources in the observed and mock (Section \ref{sec:modifications}) non-\textit{class I} samples, colored by color excess. The shaded strip corresponds to the expected location of systems with dark $0.45-0.75\,M_{\odot}$ companions. The dashed and dotted lines are the boundaries below which systems are consistent with hosting a single MS star and an inner binary, respectively (i.e. non-\textit{class I} systems lie above the dotted line). These plots can be compared to Figure 5 of \citet{Shahaf2024MNRAS}. 

In both panels, there is a high density of sources occupying the shaded strip with color excess close to zero, corresponding to true WD+MS binaries. The distribution in the mock sample is somewhat narrower and has a shallower slope than in the observed sample. This tension reflects a stronger correlation between primary and secondary masses in the mock sample than in the observed sample, as we discuss further in Section \ref{ssec:results_empirical}. 

Here, we focus on the sources with positive color excess values located outside of the strip and in between the two boundaries, which are presumably the triples. For the simulated sample, the vast majority of triples do indeed fall in this clump. However, there are significantly more such likely triples in the observed sample, likely reflecting an imperfect model of compact hierarchical triples in our simulated sample.

We could increase the total number of these sources in our simulated population by altering the mass ratio distribution from which we draw the outer tertiary's mass, resulting in fewer systems for which both stars in the inner binary have evolved into WDs (Section \ref{ssec:triples}). Still, this would fail to reproduce the pileup of observed sources at $M_1 \sim 0.95\,M_{\odot}$, also seen in Figure 6 of \citet{Shahaf2024MNRAS}. This feature may reflect additional structure in compact triple population that is not accounted for in our model. Alternatively, it could be the result of biased primary masses for the observed sources in the \texttt{binary\_masses} catalog. Since almost all triples are effectively removed by the cut on color excess, these discrepancies have a minimal effect on our final sample. 

\section{Results of E24 model} \label{sec:e24_model}

\begin{figure*}
    \centering
    \includegraphics[width=0.9\linewidth]{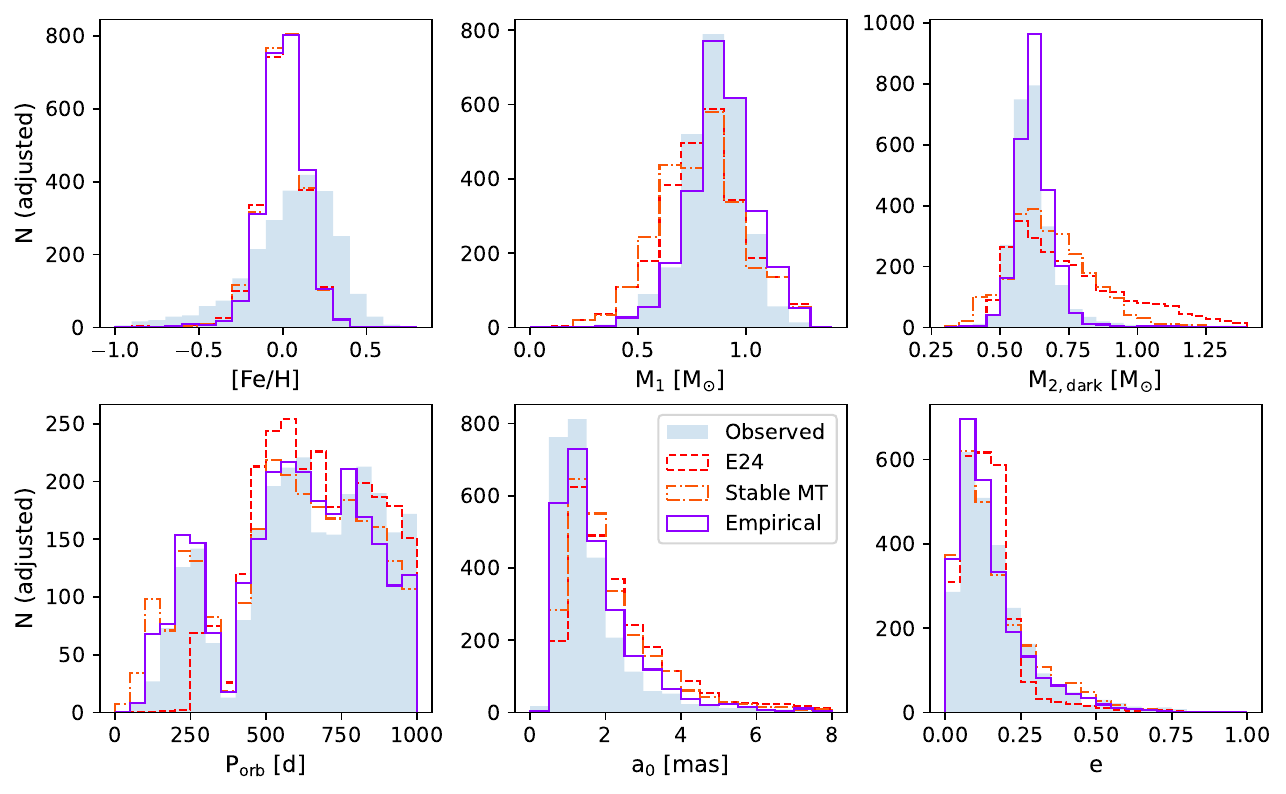}
    \caption{Distributions of stellar and orbital parameters in the observed NCE sample from \citet{Shahaf2024MNRAS}, and mock samples obtained from processing the original mock NSS catalog from \citetalias{El-Badry2024OJAp_b} (with the inclusion of hierarchical triples; Section \ref{ssec:triples}), the stable MT model (Section \ref{sssec:T23_model}), and our empirical model; Section \ref{sssec:empirical_model}). For comparison purposes, the total number of systems in all samples are made to be equal. Overall, the empirical model best reproduces the features of the observed sample.}
    \label{fig:nce_dist}
\end{figure*}

\begin{figure*}
    \centering
    \includegraphics[width=0.99\linewidth]{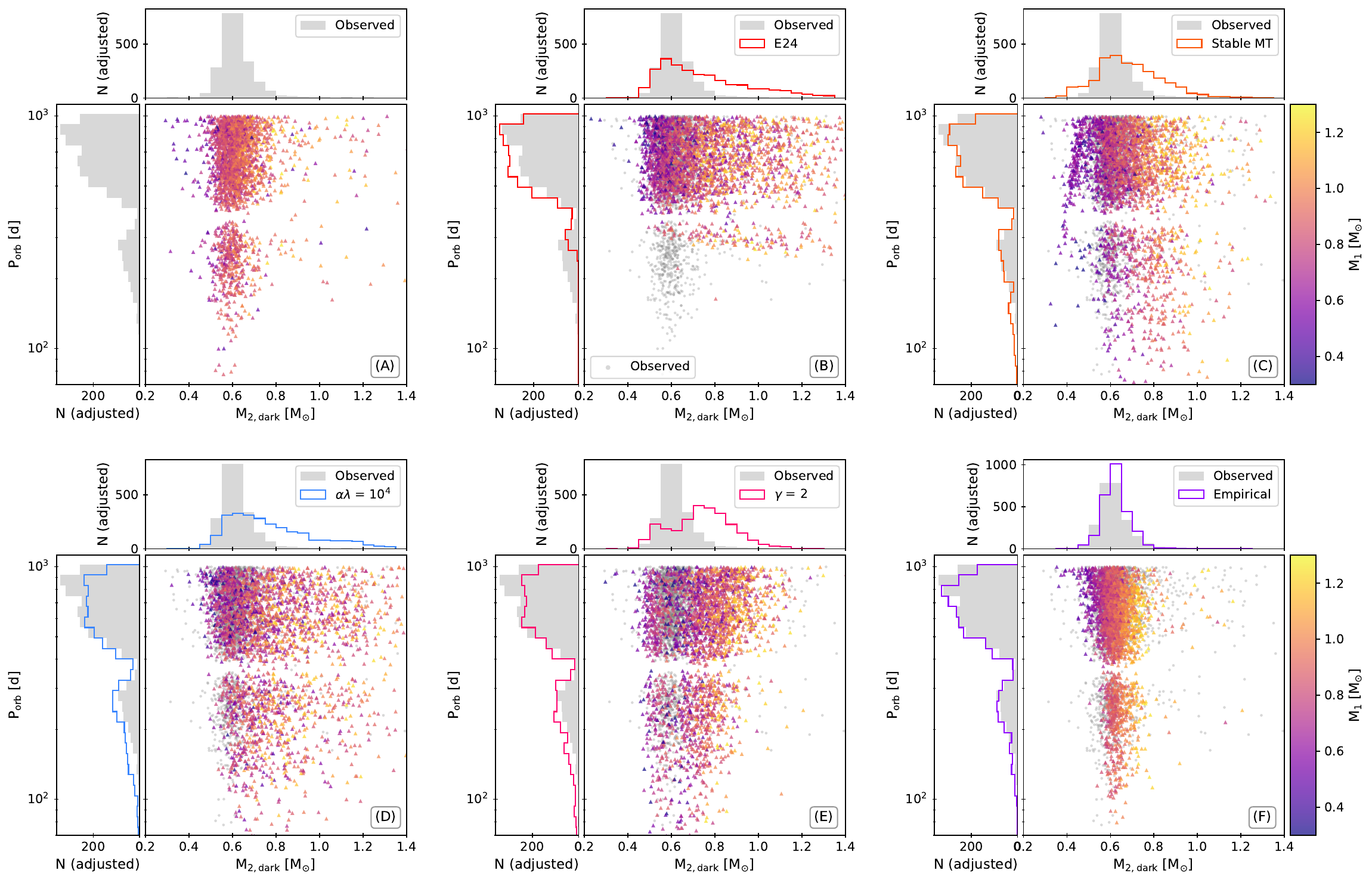}
    \caption{Orbital period and WD mass, colored by companion mass, for systems in the observed sample (Panel A; \citealt{Shahaf2024MNRAS}), the mock sample resulting from the E24 model (Panel B; Section \ref{sec:e24_model}), and various other models of the orbital evolution tested in this work (Panels C - F; Section \ref{ssec:fshrink_models}). For ease of comparison, the total number of systems in all samples are adjusted to be equal, and the observed sample is shown in gray across all panels.}
    \label{fig:porb_mwd}
\end{figure*}

We first consider the original mock NSS catalog from \citetalias{El-Badry2024OJAp_b}, with the addition of hierarchical triples (Section \ref{ssec:triples}). This will be referred to as the ``E24 model". In Figure \ref{fig:nce_dist}, we plot the distributions of several stellar and orbital parameters of the NCE sample resulting from this model (red dashed border) on top of those observed (light blue, filled). In Figure \ref{fig:porb_mwd}, we compare these samples (panels (A) and (B)) in the $P_{\rm orb}-M_{\rm 2, dark}$ plane, where $M_{\rm 2, dark}$ is the secondary mass implied by the astrometric solution assuming a completely dark companion (i.e. the approximate WD mass). The distribution of each parameter is plotted on the corresponding axis, and points are colored by companion mass.

The total number of systems in the mock sample is within $8\%$ of the true sample, suggesting that the assumed $10\%$ fraction of post-AGB MT systems in wide orbits is reasonable, given the other assumptions of the model (see discussion towards the end of Section \ref{sssec:color_excess_dist}).

We see that there is a deficit of systems with $P_{\rm orb} < 300\,$d and $e > 0.2$ in the E24 model. The color excess cut removes the vast majority of hierarchical triples that enter the non-\textit{class I} sample so that they make up less than 0.5\% of the NCE sample. Therefore, we can conclude that the deficit of systems with $e > 0.2$ cannot be attributed to contaminating triples. The distribution of color excess is discussed further for our udpated empirical model in Section \ref{sssec:color_excess_dist} (Figure \ref{fig:color_excess_dist}). The lack of shorter period systems can be addressed by modifying the conditions that lead to the formation of wide post-MT systems, as will be described in Section \ref{ssec:fshrink_models}. 

The model significantly over-predicts the fraction of WDs with masses $\gtrsim 0.8\,M_{\odot}$, which we address in our updated models as described in Section \ref{sec:modifications}. This deficit of massive WDs in the AMRF sample compared to the field WD population has been explored recently by \citet{Hallakoun2024ApJL}. Massive WDs are over-represented in the AMRF sample as they produce larger orbits and can therefore be identified as WDs even when they orbit more massive/luminous stars, so the dearth of massive WDs is highly significant.

The observed primary mass distribution is more narrow and peaks at a slightly higher mass compared to the E24 model. This suggests that there may be a relationship between the mass of the accretor in the progenitor system -- the luminous primary today -- and the probability to end up in wide orbits post-interaction that are detectable via astrometry (Section \ref{ssec:fshrink_models}).  

We also see that there is a slight discrepancy between the observed and simulated [Fe/H] distributions. This may be attributed to uncertainties in the underlying stellar population simulated by \texttt{Galaxia} or those in the measured metallicites reported by \citet{Zhang2023MNRAS} (Section \ref{sssec:color_excess_dist}). 

\section{Modifications to the E24 model} \label{sec:modifications}

We tested several modifications to the E24 model. We first describe the shared changes made across all of our new models.

\subsection{Interaction on the RGB vs. AGB} \label{ssec:Rmax_RGB_AGB}

\begin{figure}
    \centering
    \includegraphics[width=0.95\linewidth]{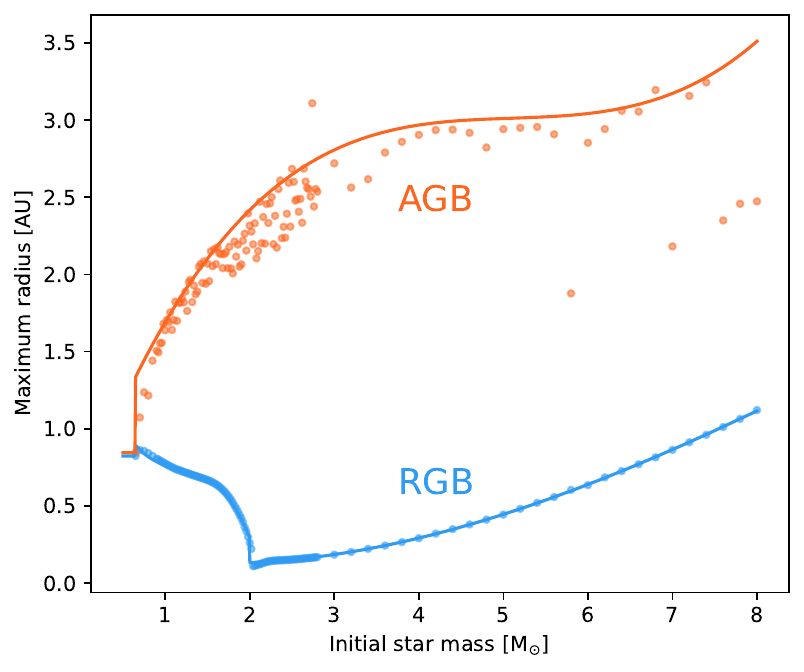}
    \caption{Maximum radius that a star reaches on the RGB and AGB as a function of its initial mass. The points are obtained from MIST evolutionary tracks, and the curves are splines fitted to the points used for interpolation. In our model, systems with initial pericenter separations below the RGB curve are assumed to become short-period PCEBs, while a fraction of those in between the two curves end up as wide PCEBs.}
    \label{fig:max_radius_mist}
\end{figure}

As mentioned in Section \ref{ssec:post_interaction_systems}, the outcome of binary mass transfer is expected to depend on the structure of the donor’s envelope. In particular, compared to red giants, AGB stars can have very loosely bound envelopes that may be ejected with very little energy \citep[e.g.][]{Belloni2024A&A, Yamaguchi2024PASP}. AGB donors may also be able to maintain stability over more extreme mass ratios \citep[e.g.][]{Temmink2023AA}. We therefore identify and separate the orbital evolution of binaries which interact when the donor is on the RGB versus the AGB by comparing their initial separations to the maximum stellar radii at each of these phases.  

We use the standard set of MIST evolutionary tracks (with rotation; \citealt{Dotter2016ApJS, Choi2016ApJ}) to predict the maximum radius reached on the RGB and AGB, $R_{\rm RGB, max}$ and $R_{\rm AGB, max}$ respectively, for a given initial stellar mass (i.e. mass of the WD progenitor). This is plotted in Figure \ref{fig:max_radius_mist}. We fit two cubic spline functions for the RGB track above and below $2\,M_{\odot}$. For the AGB track, there is large scatter due to thermal pulses, particularly above $\sim 5\,M_{\odot}$. To avoid such features, we fit the maximum value in a sliding window spanning five data points. We note that progenitors with masses above $5\,M_{\odot}$ make up a small fraction of the total population ($\sim 3\%$) and thus the effect of this choice on the statistics of the final population is small. 

Furthermore, we consider the Roche lobe radius at periastron, $r_{\rm L, peri} = f_q a_i (1-e)$ (where $f_q$ is the Eggleton factor; \citealt{Eggleton1983ApJ}, $a_i$ is the initial semi=major axis, and $e$ is the eccentricity). This is reasonable as interaction most likely begins when the two components are closest to each other. We assume that systems with $r_{\rm L, peri} < R_{\rm RGB, max}$ first interact when the WD progenitor is on the red giant branch. Meanwhile, systems with $R_{\rm RGB, max} < r_{\rm L, peri} < R_{\rm AGB, max}$ first interact with an AGB donor. In the E24 model, 10\% of the systems in the latter group were randomly chosen to end up in orbits with final separations that are a constant fraction, $f_{\rm shrink} = 0.5$, of their initial separations. We modify this evolution, testing several different models, which are described in Section \ref{ssec:fshrink_models}.

\subsection{Donor mass at the onset of MT} \label{ssec:donor_mass}

\begin{figure}
    \centering
    \includegraphics[width=0.99\linewidth]{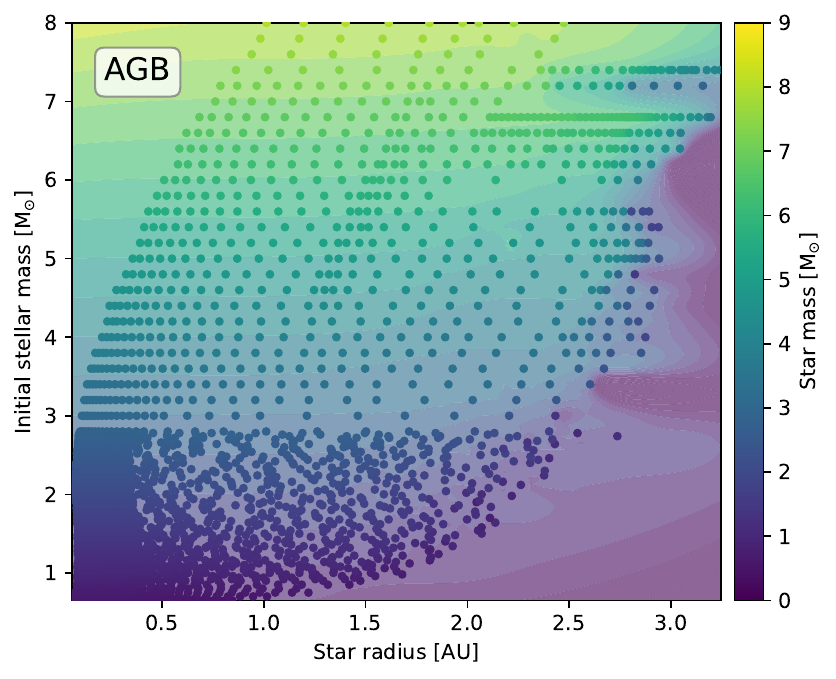}
    \caption{Stellar mass as a function of initial mass and radius on the AGB. The markers are from MIST evolutionary tracks, while the color scale shows our interpolation. We see significant mass loss occurring as the star expands to large radii near the end of the AGB phase.}
    \label{fig:stellar_mass}
\end{figure}

The models of orbital evolution described in Section \ref{ssec:fshrink_models} all depend on the donor mass, or the system mass ratio, at the onset of MT. These quantities are generally different from their initial values, as the donor may have experienced significant mass loss via winds. We determine the donor mass at Roche lobe overflow, $M_{\rm 1, MT}$, by interpolating MIST tracks to predict the stellar mass given the initial mass and radius at the time of Roche lobe overflow (Figure \ref{fig:stellar_mass}). For each system which interacts on the RGB or AGB, we use 2D interpolation with the \texttt{scipy.interpolate.RBFInterpolator} package to predict the donor's mass at the time of Roche lobe overflow. Figure \ref{fig:stellar_mass} shows the interpolated function on the AGB.  During thermal pulses, the mass-radius relation for a given initial mass can be non-monotonic. However, this is not problematic for our models, as we only consider interactions where the radius is larger than at any earlier point in the evolution (otherwise interaction would have occurred earlier).

\subsection{Eccentricity evolution} \label{ssec:eccentricity_evol}

We alter the eccentricity evolution so that the final eccentricity, $e_f$, of the wide post-MT systems are related to their initial values, $e_i$, by $e_f = 0.8e_i^4 + 0.04$. A gaussian noise with a standard deviation of 0.05 is also added, and an upper bound of $e_f = 1$ was imposed. 

This is an ad-hoc prescription which simply allows us to understand how the complicated selection function in forming the final observed sample affects its eccentricity distribution. However, simulations of CEE from \citet{Glanz2021MNRAS} found a correlation between initial and final eccentricities at the end of the dynamical phase which motivates a choice of this kind relating the two. There are also several theoretical studies of stable MT occurring in eccentric orbits, which find that orbits may not circularize as efficiently as previously assumed \citep{Hadjidemetriou1969Ap&SS, Sepinsky2007ApJ, Hamers2019ApJ, Rocha2025ApJ}. Implementing such models for the study of wide WD+MS binaries is a promising avenue for future work.

\subsection{Models for orbital evolution under MT} \label{ssec:fshrink_models}

We modify the conditions that lead to the formation of close and wide post-interaction binaries from those described in Section \ref{sec:forward_model}. We test several different models, which are described below. 

Each model of the orbital evolution produces a new simulated population and a mock NSS catalog (Section \ref{sec:forward_model}). We process each mock catalog through the \citet{Shahaf2024MNRAS} selection algorithm (Section \ref{sec:amrf_sample}) to obtain the final NCE sample of WD+MS binary candidates. Table \ref{tab:model_comparison} summarizes the features of each model and the resulting mock sample.

\begin{table*}[ht]
\centering
\begin{tabular}{l|l|p{7cm}|l|l|l|l}
\hline
                & Section & Description & Total \# & $P_{\rm orb}$ [d] & $M_{1}$ [M$_{\odot}$]  & $M_{\rm 2, dark}$ [M$_{\odot}$] \\ \hline\hline
Observed        & \ref{sec:amrf_sample} & Sample of Gaia WD+MS binary candidates from \citet{Shahaf2024MNRAS}. & 3145   &  $640\pm234$   &  $0.86\pm0.14$    &  $0.61\pm0.12$   \\ \hline
E24       & \ref{sec:forward_model} & 10\% of post-AGB MT orbits remain wide at half their initial separations. All other post-MT binaries end up in $1\,$d periods. & 2900  &  $663 \pm 184$   &  $0.80\pm0.20$  &  $0.70 \pm 0.21$  \\ 
Stable MT & \ref{sssec:T23_model} & Use \citet{Temmink2023AA} quasi-adiabatic criterion for critical mass ratios. For stable MT, evolve orbits according to \citet{Soberman1997AA} with $\beta = 1$. For systems with RGB donors, calculate WD masses so that final periods are consistent with relation from \citet{Rappaport1995MNRAS}. All other post-MT binaries end up in $1\,$d periods. &  9900  & $602 \pm 245$   & $0.80 \pm 0.20$ &   $0.67 \pm  0.15$ \\ 
$\alpha\lambda$ & \ref{sssec:alpha_formalism} & Separations of all post-AGB MT orbits determined by the $\alpha$-formalism, with $\alpha\lambda = 10^4$. All other post-MT binaries end up in $1\,$d periods. & 31000  &  $543 \pm 252$   & $0.82 \pm 0.19$   & $0.72 \pm 0.19$   \\ 
$\gamma$  & \ref{sssec:gamma_formalism} & Separations of all post-AGB MT orbits determined by the $\gamma$-formalism, with $\gamma = 2$. All other post-MT binaries end up in $1\,$d periods. & 17000 & $554 \pm 260$   &  $0.81 \pm 0.20$  & $0.74 \pm 0.14$   \\ 
Empirical  & \ref{sssec:empirical_model} & Fixed critical mass ratio of 0.38. For stable MT from AGB donors, evolve orbits according to \citet{Soberman1997AA} with $\beta = 1$. All other post-MT binaries end up in $1\,$d periods. & 3100  &  $609 \pm 234$  & $0.89 \pm 0.15$   & $0.62 \pm 0.06$  \\ \hline
\end{tabular}
\caption{Summary of the observed NCE sample and simulated NCE samples resulting from different models of the orbital evolution due to MT. Total numbers are rounded to 2 significant figures. The three rightmost columns list the median and standard deviation of each parameter within each sample.}
\label{tab:model_comparison}
\end{table*}

\subsubsection{Model for stable MT} \label{sssec:T23_model}

First, we describe a model which assumes that the AU-scale WD+MS binaries are post-stable MT systems. In this case, we need to determine whether each interacting system undergoes stable or unstable MT.  \citet{Temmink2023AA} ran a large grid of 1D binary models and determined the ``quasi-adiabatic" stability boundary. The corresponding critical accretor-to-donor mass ratio above which MT remains stable, $q_{\rm qad}$, depends on the mass and radius of the donor (Figure 8 of \citealt{Temmink2023AA}). 

We determine the orbital evolution due to stable MT using relations derived by \citet{Soberman1997AA}. We assume fully non-conservative mass transfer with isotropic re-emission from the accretor (i.e. $\beta = 1.0$). In this limit, the ratio of the final to initial separation, $a_f/a_i$, is given by: 
\begin{equation} \label{eqn:fshrink_S97}
    \frac{a_f}{a_i} = \left( \frac{q_{\rm f}}{q_{\rm MT}} \right)^{3} \left( \frac{1 + q_{\rm f}}{1 + q_{\rm MT}} \right) e^{2 (1/q_{\rm f} - 1/q_{\rm MT})}
\end{equation} 
where $q_{\rm f} = M_{2, \rm init}/M_{\rm WD}$ (\citealt{Soberman1997AA} use donor-to-accretor mass ratios, which we have inverted here). If $q_{\rm f}$ is sufficiently large compared to $q_{\rm MT}$, $a_f/a_i > 1$, and the orbit widens. 

\citet{Temmink2023AA} consider fully conservative MT. Typically, non-conservative MT is expected to lead to increased stability, so the critical mass ratios used can be regarded as upper limits. Accordingly, our modeling may be expected to under-predict the number of stable MT products. In fact, we show in Section \ref{ssec:results_T23} that the model significantly {\it over}-predicts the number of detectable systems. 

Additionally, \citet{Temmink2023AA} do not model AGB donors with masses below $2\,M_{\odot}$, as they find that stars in this mass range do not expand significantly beyond the tip of the RGB until the start of thermal pulses, where \citet{Temmink2023AA} end their simulations. We find that AGB donors less massive than $2\,M_{\odot}$ make up a sizable fraction ($\sim 30\%$) of all AGB donors. By default, we assume that all of these systems undergo stable MT. For donors above $2\,M_{\odot}$, we compare the mass ratio at the onset of MT to $q_{\rm qad}$ obtained by interpolating tabulated values from \citet{Temmink2023AA}.

It has long been appreciated that many low-mass WDs formed through stable MT from RGB donors lie on a tight relation between period and WD mass \citep[e.g.][]{Refsdal1971A&A, Chen2013MNRAS, Istrate2014A&A, El-Badry2022MNRAS}.  \citet{Rappaport1995MNRAS} derived a fitting formula for this relation which remains widely used today. Therefore, for donors that overflow their Roche lobes on the RGB, we solve for final WD masses by equating the relation from \citet{Rappaport1995MNRAS} to Equation \ref{eqn:fshrink_S97}. While \citet{Rappaport1995MNRAS} only considered donors with masses up to $2\,M_{\odot}$, this is not a problem here, as MT with more massive RGB donors become unstable in the vast majority of cases. For the few exceptions, we continue using the isolated WD IFMR from \citet{Weidemann2000AA}, which is our default choice unless stated otherwise. 

For systems that interact with AGB donors, we also adopt the \citet{Weidemann2000AA} IFMR. However, if the core growth is effectively terminated by the onset of MT, we might expect less massive WDs for a given initial stellar mass. An alternative approach would then be to interpolate over a grid of stellar radii and masses to estimate the core mass at Roche lobe overflow (analogous to how donor masses were obtained in Section \ref{ssec:donor_mass}). We test this in Appendix \ref{ssec:appendix_mod_wd_ifmr} and find no significant improvement in the match between the resulting mock and observed NCE samples, so we choose not to truncate core growth in the default model. 

All interacting systems which experience unstable MT (i.e. CEE) are assumed to end up in short ($1\,$d) periods and circular orbits. We make the same assumption for cases where MT begins before the donor is on the RGB (i.e. Hertzsprung gap/sub-giant donors). This is inconsequential to our results as such systems produce binaries containing very low-mass WDs with short orbital periods, which are unlikely to make it into the NSS catalog and make it past the AMRF cuts. 

\begin{figure*}
    \centering
    \includegraphics[width=0.9\linewidth]{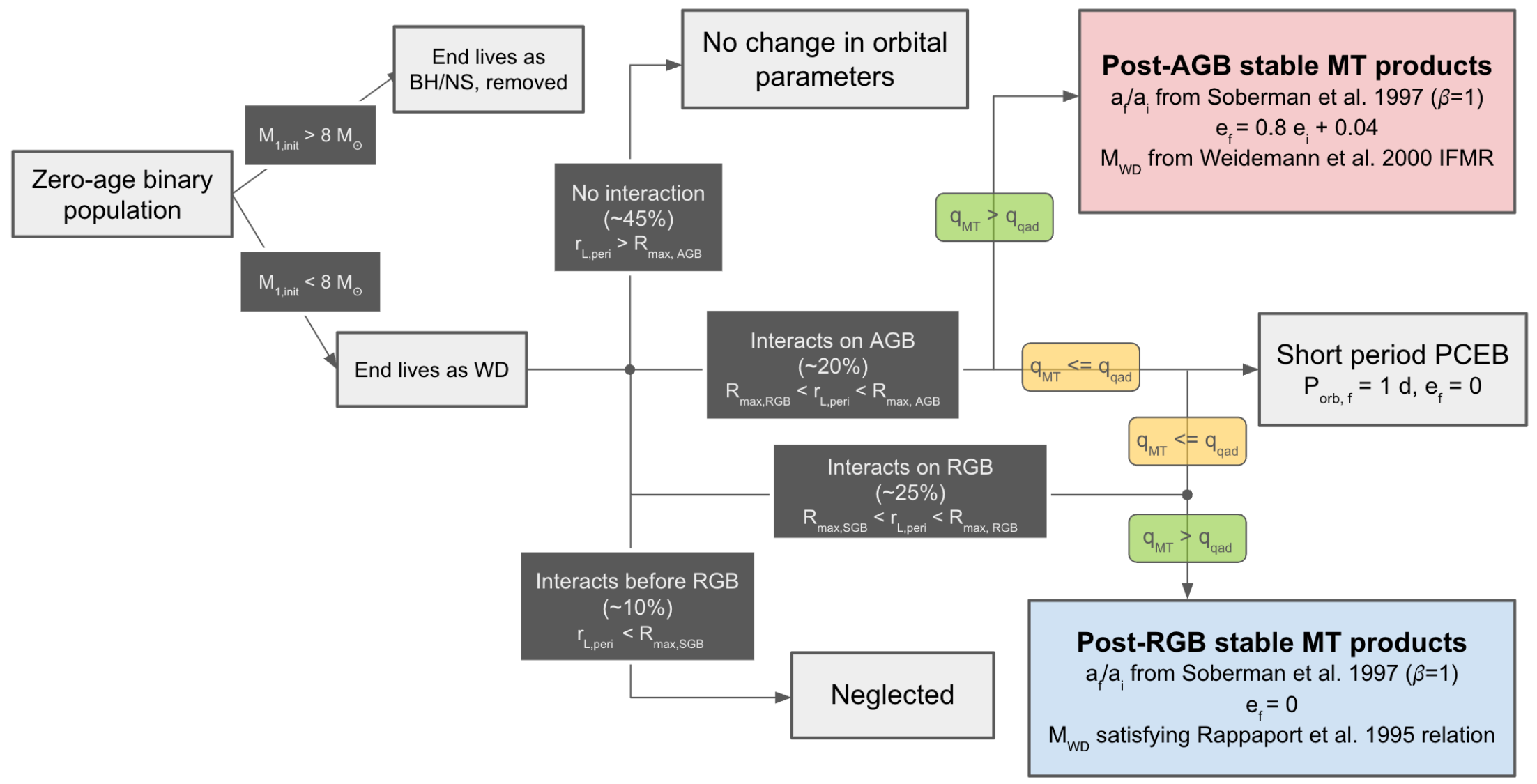}
    \caption{Schematic summarizing the assumed orbital evolution of the zero-age binary population in the stable MT model, which leads to the formation of short-period and wide post-MT binaries. Detailed descriptions of the components can be found in Section \ref{sssec:T23_model}.}
    \label{fig:schematic_T23_S97}
\end{figure*}

Moving forward, these assumptions will be referred to as the ``stable MT model''. The key components of the model are illustrated in Figure \ref{fig:schematic_T23_S97}.

\subsubsection{An $\alpha$-formalism model} \label{sssec:alpha_formalism}

It is uncertain whether stable mass transfer or CEE is a more accurate description of the binary interaction that leads to the formation of AU-scale post-MT binaries. Therefore, we also run a model using the $\alpha$-formalism prescription \citep{Livio1988ApJ, deKool1990ApJ}, often used to describe orbital evolution under CEE \citep[e.g.][]{Zorotovic2010A&A, Davis2012MNRAS, Scherbak2023MNRAS}.

Here, we test whether the $\alpha$-formalism alone can predict the properties of WD+MS binaries formed via interaction on the AGB. At its core, the formalism is a statement of energy conservation: 
\begin{equation} 
    E_{\rm bind} = \alpha\left(-\frac{GM_{\rm WD}M_{\star}}{2a_f} + \frac{GM_{i}M_{\star}}{2a_i} \right)
\end{equation} 
where $E_{\rm bind}$ is the binding energy of the donor's envelope, and the right-hand side is a fraction $\alpha$ of the orbital energy loss. $M_{i} = M_{\rm 1, MT}$ is the initial mass of the WD progenitor (i.e. the donor). In the classic formulation, $E_{\rm bind} = -(GM_iM_{i, \rm env})/(\lambda R_i)$ where $M_{i,\rm env}$ and $R_i$ are the envelope mass and radius of the donor, and $\lambda$ is a free parameter which encodes information about the envelope structure, often set to $\sim 0.5$. Taking $R_i = r_{\rm L, peri}$, the Roche lobe radius at periastron, nd $M_{i, \rm env} = M_i - M_{\rm WD}$, we can solve for $a_f$ with one free parameter, namely the product $\alpha\lambda$: 
\begin{equation} \label{eqn:alpha_lambda}
    \frac{a_f}{a_i} = \frac{M_{\rm WD}}{M_i}\left(1 + \frac{2M_{i, env}a_i}{M_{\star}R_{i}\alpha\lambda} \right)^{-1}
\end{equation} 
We apply this to all systems that experience MT from AGB donors. All other systems which interact with RGB donors are assumed to end up in $1\,$d circular orbits. 

A value of $\alpha\lambda = 1$ is typical under the standard assumptions because $\alpha$ is defined to be less than or equal to 1 and considering only the gravitational binding energy, we expect $\lambda$ to be on the order unity. A higher value may be understood as the donor having a particularly loosely bound envelope, which may be the case if it is undergoing a thermal pulse on the AGB, or if additional sources of energy (e.g. recombination) aid in the unbinding \citep[e.g.][]{Ivanova2013A&ARv, Ivanova2015MNRAS, Belloni2024A&A, Yamaguchi2024PASP, Yamaguchi2024MNRAS}. Several earlier works have found large (even negative, considering recombination) values of $\lambda$ for AGB stars \citep[e.g.][]{Paczynski1968AcA, Han1994MNRAS, Dewi2000A&A}.

\begin{figure}
    \centering
    \includegraphics[width=0.97\linewidth]{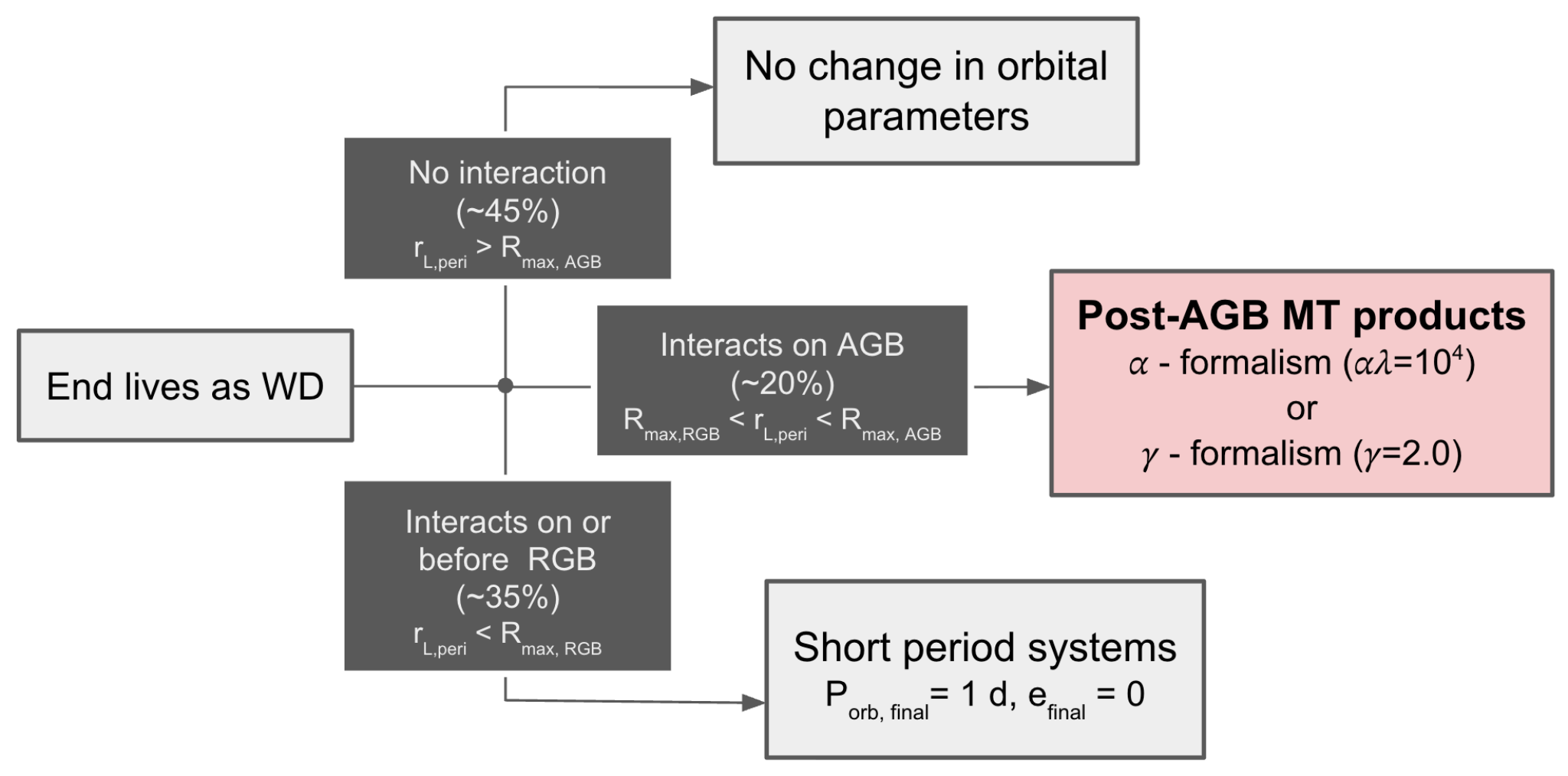}
    \caption{Components of the $\alpha\lambda$ and $\gamma$ models distinct from the stable MT model illustrated in Figure \ref{fig:schematic_T23_S97}. Detailed descriptions are provided in Sections \ref{sssec:alpha_formalism} and \ref{sssec:gamma_formalism}.}
    \label{fig:schematic_2}
\end{figure}

We will refer to this model as the ``$\alpha\lambda$ model" hereafter. Although the $\alpha$-formalism is not expected to provide an accurate description of the mass transfer process, we include this model as it may be useful to have an effective value of $\alpha\lambda$ to predict properties of PCEBs without having to use detailed evolutionary models, which themselves struggle to predict the outcomes of CEE. The results of using this model can be found in Section \ref{ssec:results_alpha_lambda} and its components are illustrated in Figure \ref{fig:schematic_2}.

\subsubsection{A $\gamma$-formalism model} \label{sssec:gamma_formalism}

We implement another commonly used prescription in the CEE literature, the $\gamma$-formalism. The $\gamma$-formalism was originally introduced to describe the first MT phase in the formation of double helium WD binaries (He DWDs). It is derived from angular momentum balance and may be appropriate if the common envelope is brought into co-rotation with the orbit such that there are no drag forces and thus little orbital shrinkage \citep{Nelemans2000A&A, Nelemans2005MNRAS}. Indeed, the orbit can \textit{widen} under this formalism, and the change in separation is given by:
\begin{equation} \label{eqn:fshrink_gamma}
   \frac{a_f}{a_i} = \left(\frac{M_i}{M_{\rm WD} }\right)^{2} \left( \frac{M_{\rm WD} + M_{\star}}{M_i + M_{\star}} \right) \left(1-\gamma \frac{M_{\rm env}}{M_i + M_{\star}} \right)^2 
\end{equation} 
where the parameters are the same as in equation \ref{eqn:alpha_lambda}. We only apply Equation~\ref{eqn:fshrink_gamma} for systems with AGB donors, and we adopt $\gamma=2$ as the fiducial value. 

Physically, $\gamma$ may be thought of as describing the average amount of specific angular momentum carried away by mass lost from the system. Although only a narrow range of $\gamma \sim 1.5 - 1.7$ has been shown to be consistent with all He DWDs \citep{Nelemans2000A&A, Nelemans2005MNRAS}, the constraining power of this parameter has since been called into question \citep{Webbink2008ASSL, Woods2011ASPC, Ivanova2013A&ARv}. The $\gamma$ formalism has also been interpreted to be a  description of stable MT as opposed to the CEE \citep{Ivanova2013A&ARv, Scherbak2023MNRAS}. Nevertheless, since this model remains commonly used in the literature, we apply it to the WD+MS sample (``$\gamma$ model") and present results in Section \ref{ssec:results_gamma}.

We also mention here the `grazing envelope evolution' (GEE; \citealt{Soker2015ApJ}) where steady jets from the donor launches material at the outskirts of the accretor's envelope. This can lead to the system avoiding the CEE and forming wide, eccentric orbits \citep{Kashi2018MNRAS, Abu-Backer2018ApJ}. While work remains to define the parameter space over which GEE is expected to occur and model the process with 3D hydrodynamical simulations, this is another possible scenario to explain the intermediate post-AGB systems and is an additional avenue to test in the future. 

\subsubsection{Fixed critical mass ratio} \label{sssec:empirical_model}

As will be described in Section \ref{sec:results_modified}, we find that all physically motivated models described above fail to reproduce some feature of the observed WD+MS binary sample. To better match the population demographics of the observed sample and infer the intrinsic population of post-MT binaries, we construct an ``empirical model'' that is tuned to produce an AMRF sample that matches the observed one.

For this model, we set a fixed value for the critical mass ratio, $q_{\rm crit}$ at the onset of MT from AGB donors. For $q>q_{\rm crit}$, we implement the \citet{Soberman1997AA} relation for $a_f/a_i$, as in the stable MT model (Equation \ref{eqn:fshrink_S97}). We ultimately choose a critical accretor-to-donor mass ratio of $q_{\rm crit} = 0.38$, which leads to a final WD+MS binary sample similar in size to that observed (Section \ref{ssec:results_empirical}). This results in about $5\%$ of post-AGB interaction systems having orbits with final periods of $P_{\rm orb} = 100-1000\,$d. Although $\sim 63\%$ of interacting systems in our model have $q > q_{\rm crit}$, the majority end up with $P_{\rm orb}>1000\,$d, and will thus only be accessible in future Gaia data releases.

In the empirical model, we assume that all binaries with RGB or Hertzsprung-gap donors donors shrink to small orbits with $1\,$d periods, preventing them from entering the NSS catalog. We make this choice because in the observed sample, there is no group of systems with periods and WD masses that lie on the expected relation for formation through stable MT from RGB donors (\citealt{Rappaport1995MNRAS}; Section \ref{sec:results_modified}).

It is natural to interpret the empirical model as a measurement of the average stability boundary for AGB donors. However, we emphasize that the adopted value of $q_{\rm crit}=0.38$ is purely empirical, chosen only to reproduce the observed mass and period distributions. The key input here is simply that binaries with more extreme mass ratios shrink to small orbits, whereas those with closer-to-equal mass ratios remain relatively wide. In reality, $q_{\rm crit}$ is likely to depend on the state of the donor. 

\section{Results of modified models} \label{sec:results_modified}

\subsection{Stable MT model} \label{ssec:results_T23}

We begin with results of the stable MT model, which makes use of the critical mass ratios from \citet{Temmink2023AA} and orbital evolution from \citet{Soberman1997AA} (Section \ref{sssec:T23_model}). The resulting mock NCE sample is shown in  $P_{\rm orb} - M_{\rm 2, dark}$ space in panel C of Figure \ref{fig:porb_mwd}. Distributions of several other parameters are shown in Figure \ref{fig:nce_dist} with an orange dashdot border. 

The total number of systems in this mock sample is approximately 3 times greater than observed, suggesting that there are too many stable MT products which remain in sufficiently wide orbits for detection. This is despite the fact that the critical mass ratios from \citet{Temmink2023AA} assumes fully conservative MT, which is expected to represent a lower limit on the fraction of MT episodes that remain stable. While we assume fully non-conservative MT in the default model, we note that the discrepancy persists if we instead assume fully conservative MT (i.e. $\beta=0$ and $\epsilon = 1$ in the notation of \citealt{Soberman1997AA}). The tension also  persists even if we assume that all systems with AGB donors less massive than $2\,M_{\odot}$ experience CEE and shrink to small orbits, in which case the total number of WD+MS binaries in the final sample over-predicted by a factor of $\sim 2.5$. 

The stable MT model also predicts a strip of systems hosting low-mass WDs below $\sim 0.55\,M_{\odot}$ that is not present in the observed sample. These are products of stable MT on the RGB and lie on the period-WD mass relation from \citet{Rappaport1995MNRAS}. The fact that these systems are not observed is another sign that the critical mass ratios from \citet{Temmink2023AA} may be too extreme, predicting MT to be more stable than it is in nature. 

Compared to the E24 sample, there are more systems with $P_{\rm orb}\lesssim 1\,$yr, in better agreement with the observed sample. This is primarily a result of the average value of $a_f/a_i$ calculated using Equation \ref{eqn:fshrink_S97} being greater than the 0.5 used in the E24 model. This means that while there are fewer systems in initially wide orbits that interact and shrink to periods short enough to be constrained in DR3, there are more systems in initially smaller orbits that remain wide enough to be detected. The adoption of more realistic boundaries for MT from RGB vs. AGB donors (Section \ref{ssec:Rmax_RGB_AGB}) also affects which systems ultimately end up in detectable orbits. 

The stable MT model also predicts fewer high-mass ($\gtrsim 1.0M_{\odot}$) WDs compared to the E24 model. This is because massive donors, which form massive WDs, tend to have more unequal mass ratios when MT commences, making mass transfer less likely to remain stable according to \citet{Temmink2023AA}. This was not modeled in the E24 model, where the outcome of binary interactions is independent of the mass ratio.

Still, this model significantly overpredicts the number of WDs above $\sim 0.7M_{\odot}$ compared to the observed sample -- another hint that its stability criteria may be too loose. In addition, the primary mass distribution is similar to that of the E24 model, being broader and peaked at a lower value than observed.

By construction, the eccentricity distributions of our updated models all match the observed sample. We emphasize that our simulations self-consistently account for eccentricity bias in the astrometric orbits \citep[e.g.][]{Wu2024arXiv}, and still we require an underlying eccentricity distribution that peaks around $0.1$ to reproduce the observed eccentricities of WD+MS binaries in astrometric orbits. While some previous works have suggested that these non-zero eccentricities are a result of triple dynamics \citep[e.g.][]{Belloni2024A&A}, we consider it unlikely that {\it all} eccentric binaries -- a large majority of the sample -- have unseen tertiaries.

\subsection{$\alpha\lambda$ model} \label{ssec:results_alpha_lambda}

Next, we discuss predictions of the $\alpha\lambda$ model (Section \ref{sssec:alpha_formalism}), where the $\alpha$-formalism prescription is used to describe the orbital shrinkage resulting from a common envelope phase with an AGB donor. 

Firstly, setting $\alpha\lambda = 1$ corresponds to $f_{\rm shrink} \lesssim 0.1$. This results in very few systems entering the NSS catalog and ultimately, the final NCE sample. This issue has already been explored by several recent works \citep[e.g.][]{Yamaguchi2024PASP, Yamaguchi2024MNRAS, Belloni2024A&A} -- the AU-scale WD+MS binaries discovered by Gaia cannot be explained as a result of CEE with $\alpha\lambda \sim 1$  \citep{Zorotovic2010A&A, Davis2012MNRAS, Toonen2013A&A, Scherbak2023MNRAS}. Instead, they require for the donor's envelope to be almost unbound. 

In Panel D of  Figure \ref{fig:porb_mwd}, we show the results of setting $\alpha\lambda = 10^4$ on the $P_{\rm orb}-M_{\rm WD}$ space, corresponding to a range of $f_{\rm shrink} \sim 0.15 - 0.65$. In this case, the model predicts about 10 times more WD+MS binaries in the final sample than are observed. To remedy this tension, we can assume that only 10\% of systems that interact on the AGB have $\alpha \lambda = 10^4$, while the remainder have $\alpha\lambda \sim 1$ and form short-period PCEBs. Still, the model predicts proportionally too many systems with $P_{\rm orb} \lesssim 1\,$yr and too few with $P_{\rm orb} \gtrsim 750\,$d. This is because the intrinsic period distribution of WD+MS binaries produced by this model rises more rapidly towards shorter periods compared to the stable MT or empirical models (Figure \ref{fig:wdms_binaries_all}). Further increasing $\alpha\lambda$ does not alleviate this problem as $a_f/a_i$ asymptotes to the ratio of the WD mass over its progenitor mass as $\alpha\lambda \rightarrow \infty$ (Equation \ref{eqn:alpha_lambda}). Moreover, this model does not reproduce the highly peaked WD mass distribution. This is because at sufficiently large values of $\alpha\lambda$, $a_f/a_i$ depends weakly on the initial mass ratio of the system. Given these limitations, we conclude that the $\alpha$-formalism is not well suited to describe the formation of these wide WD+MS binaries. All of the same issues exist for a more modest value of $\alpha\lambda = 100$, but with an even more disproportionate abundance of systems with $P_{\rm orb} \lesssim 1\,$yr.  

\subsection{$\gamma$ model} \label{ssec:results_gamma}

We present predictions of the $\gamma$ model (Section \ref{sssec:gamma_formalism}), where the $\gamma$-formalism was used to predict the amount of orbital shrinkage that occurs due to MT from an AGB donor. 

As shown in Panel E of Figure \ref{fig:porb_mwd}, we find that the predicted population suffers many of the same problems as the population predicted by the $\alpha\lambda$ model, though less severely. We find that $\gamma = 2.0$ results in slightly more systems at the longest periods compared to the $\alpha\lambda$ model. A lower value of $\gamma = 1.5$, as inferred for He DWDs, significantly overproduces the number of systems at $P_{\rm orb} \lesssim 1\,$yr. It is unsurprising that the model predicts long periods, as the $\gamma$-formalism was introduced precisely to allow for wider post-MT separations that could be not be produced by $\alpha\lambda$ model. 

We see that the WD mass distribution is double peaked with a trough at $\sim 0.6\,M_{\odot}$. This is a consequence of the behavior of Equation \ref{eqn:fshrink_gamma}, which has a non-monotonic relation of the WD mass with a local minimum around this value for $\gamma = 2.0$. 

Moreover, we find the primary mass distribution remains broad compared to the observed sample. This is because for values of $a_f/a_i$ such that the final separation is detectable with astrometry, there is a weak dependence on the accretor (and ultimately, the final primary) mass. Moreover, for a single value of $\gamma = 2.0$, there is about five times more systems that enter the final WD+MS binary sample than observed. 

\subsection{Empirical model} \label{ssec:results_empirical}

Lastly, the distributions of stellar and orbital parameters of the mock NCE sample from our empirical model are plotted with a solid purple border on Figure \ref{fig:nce_dist}. In Panel F of Figure \ref{fig:porb_mwd}, we also plot the systems in $P_{\rm orb}-M_{\rm WD}$ space. 

Compared to the E24 and stable MT models, this model leads to a primary mass distribution that is narrower and peaked at a higher value, and a WD mass distribution with a stronger deficit above $0.75\,M_{\odot}$, in good agreement with the observed distributions. This occurs because in the empirical model, the MT stability is a step function in $q$, such that only systems with initial  mass ratios above $0.38$ remain wide. This is a stricter cut than the stable MT model, where stability is maintained for a wider range of mass ratios. In this scheme, less massive donors with more massive companions are strongly favored to form wide post-MT systems that are detectable by astrometry. This also alleviates the discrepancy in the total number of systems in the final NCE sample found for the stable MT model, producing a final NCE sample with almost exactly the same number of systems as observed.  

By construction, there is no strip of stable MT products from RGB donors predicted for the empirical model. This suggests that the majority of systems that interact with RGB donors undergo CEE and shrink to short orbital periods. The period distribution of the empirical model matches the observed distribution reasonably well.  

Comparing the color gradient of the data points in Panels A and F of Figure \ref{fig:porb_mwd}, we note that that the correlation between the WD and MS star masses is stronger in the simulated sample than in the observed sample. A correlation between the two is expected, as systems hosting more massive MS stars must be accompanied by a more massive WD for them to make it past the AMRF cut. This leads to all systems in both the observed and simulated samples having WD-to-MS mass ratios above $\sim0.6$. However, the spread above this lower bound is consistently narrower for the simulated sample, across all models. This is also reflected in the different slopes of the dense strip of points corresponding to the WD+MS binaries on the AMRF-$M_1$ space, as seen in Figure \ref{fig:amrf_m1}. This suggests that the empirical model does not capture all the physics relevant to setting the WD or primary masses, perhaps due to truncation of AGB stars' core growth after MT begins. 

\subsection{Intrinsic properties of WD+MS binaries} 

In the previous section, we made changes to the simulated binary population which was then processed by the NSS \citep{Halbwachs2023A&A} and AMRF \citep{Shahaf2024MNRAS} analysis pipelines to produce a mock NCE sample whose parameter distributions approximately match the Gaia WD+MS binary sample. We are now in a position to study the properties of the underlying WD+MS binary population. As our empirical model best reproduced the features of the \citet{Shahaf2024MNRAS} sample (Section \ref{ssec:results_empirical}), we discuss the simulated population using this model here. 

\subsubsection{Completeness fractions} \label{sssec:completeness_fracs}

\begin{figure*}
    \centering
    \includegraphics[width=0.9\linewidth]{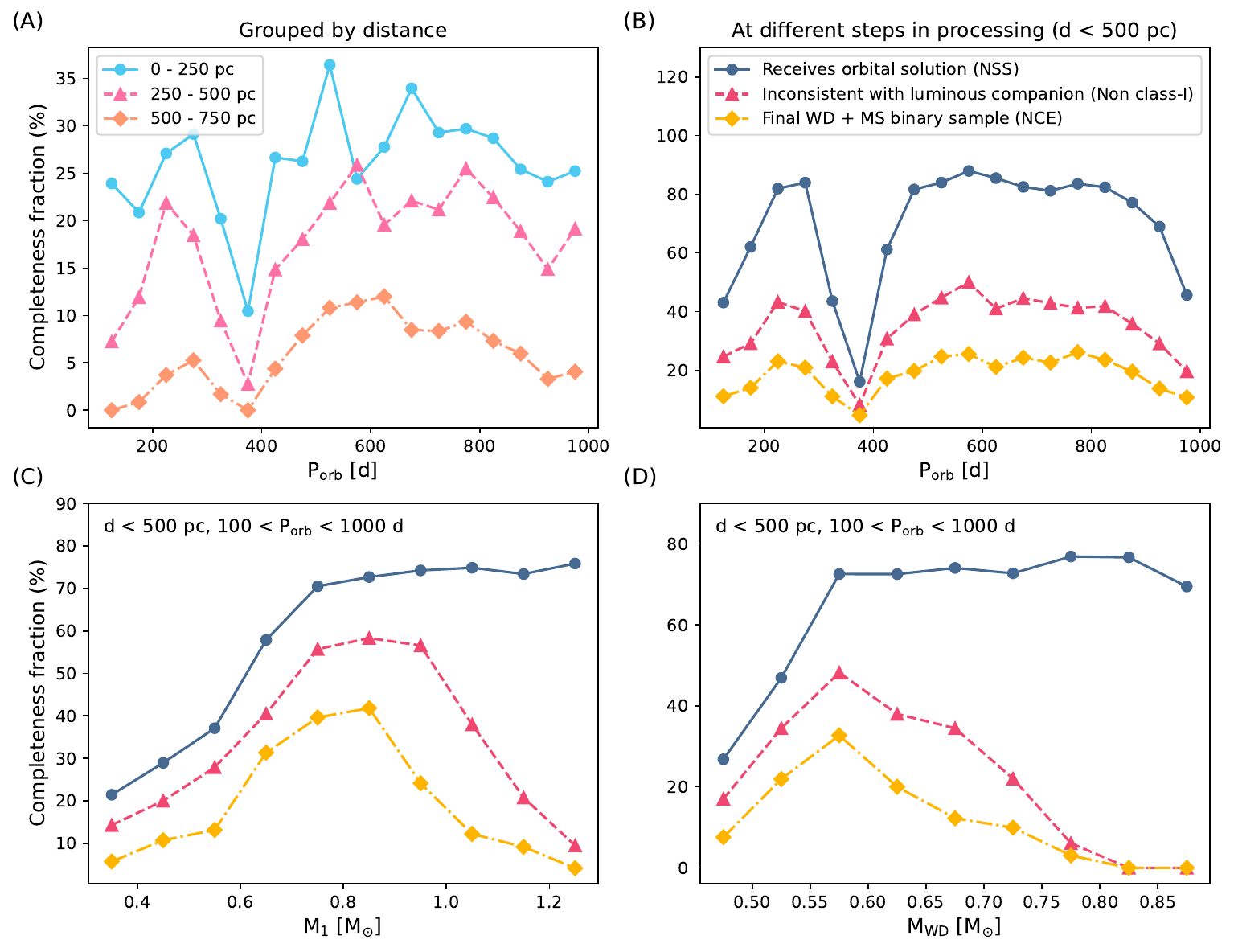}
    \caption{(\textit{A}): Completeness fraction of the mock NCE sample binned by orbital period and grouped by distance ranges, for our empirical model. As expected, the completeness falls at further distances. The dip in completeness at $\sim 1\,$yr is due to the confusion between parallactic and orbital motion. (\textit{B}): Completeness fraction binned by orbital period at different stages in the processing pipeline for sources within $ 500\,$pc. (\textit{C}): Same as panel B, but binned by primary mass. We see that while completeness for the NSS catalog generally rises for more massive primaries as they are brighter on average, they are disfavored by the cut on AMRF. (\textit{D}): Same as panel B, but binned by WD mass. While there is a bias towards more massive WDs for a given primary mass for all processing steps, the positive correlation between the two complicates this.}
    \label{fig:completeness}
\end{figure*}

We first calculate the completeness fraction as a function of a few different parameters. We define completeness as the number of systems that make it into the final NCE sample in each parameter bin divided by the total number of all WD+MS binaries in the simulated population in the same bin. For the orbital period, we take bins of width $50\,$d from $100$ to $1000\,$d, and consider sources in several distance ranges. The resulting completeness fractions are plotted for our empirical model in panel A of Figure \ref{fig:completeness}. In panels B, C, and D we plot the completeness fractions at different steps in the processing pipeline (for sources at distances $< 500\,$pc) binned by period, primary mass, and WD mass, respectively. 

The dip at $\sim 1\,$yr is a selection effect that results from confusion between parallactic and orbital motion. This prevents systems with orbital periods close to 1 yr from entering the NSS catalog, and any of its products (including the non-\textit{class I} and NCE samples). As expected, the completeness drops at all periods as we move out to larger distances. The total number of systems in the simulated population is greatest below $50\,$d but due to their small orbits, almost none of them receive orbital solutions in the NSS catalog, making their completeness close to zero. 

Moreover, we see that completeness of the NSS catalog increases with primary mass as more massive primaries are, on average, more luminous. Meanwhile, there is a sharp drop off beyond $\sim 0.8\,M_{\odot}$ for the non-\textit{class I} sample, since the AMRF cuts disfavor massive primaries. In addition, more massive primaries evolve faster and are thus redder on average, making them more likely to be removed by the cut on color excess. 

Lastly, we find that the completeness is generally higher for more massive WDs in the NSS catalog as they will produce wider orbits at fixed period, which are easier to detect. The effect of the AMRF cut on the completeness is more complicated, because while more massive WDs are favored for a given primary mass, they are more likely to be found around more massive and luminous primaries. These trends are described further in Section \ref{ssec:parameter_dist}. 

\subsubsection{Parameter distributions} \label{ssec:parameter_dist}

\begin{figure*}
    \centering
    \includegraphics[width=0.99\linewidth]{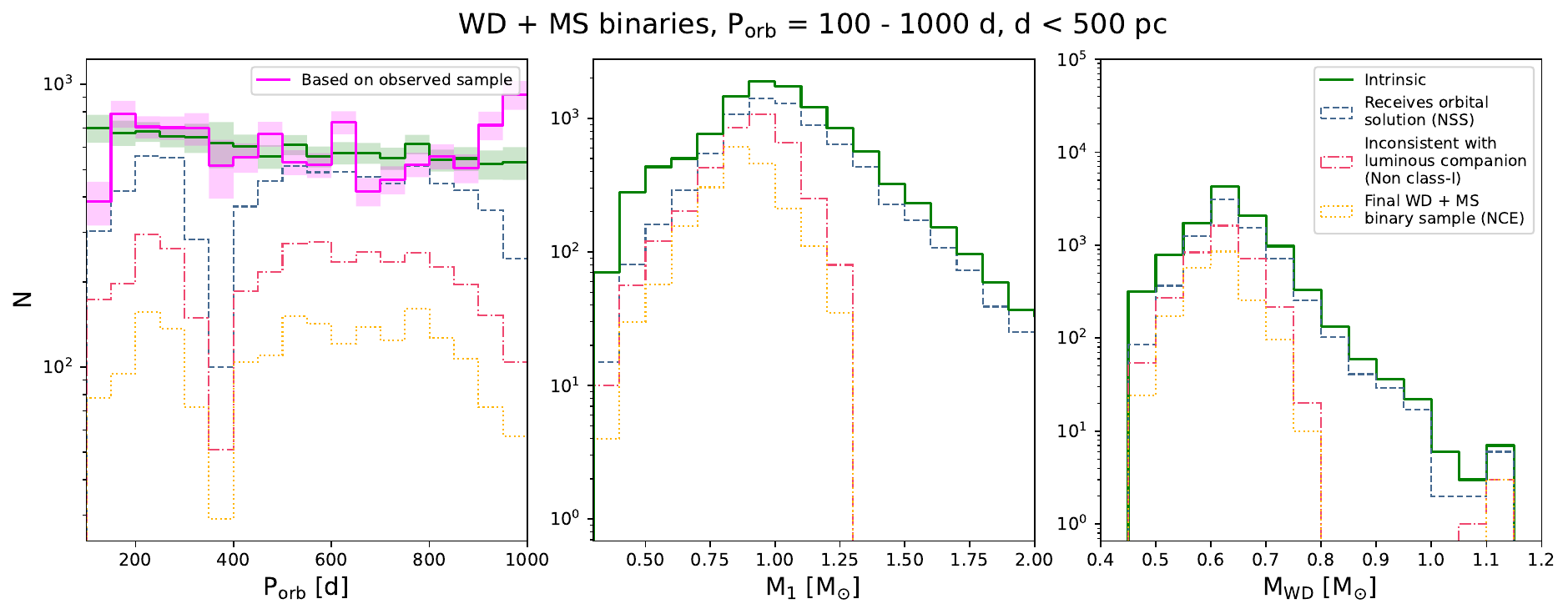}
    \caption{Distributions of orbital period and component masses of WD+MS binaries, with periods between $100-1000\,$d and at distances within 500 pc, at different stages in our analysis: the full simulated WD+MS binary population, systems that enter the mock NSS catalog, those in the non-\textit{class-I} sample, and lastly, those in the NCE sample. In magenta, we have the intrinsic period distribution inferred by taking the distribution of the observed NCE sample and correcting it with the completeness implied from our models (Figure \ref{fig:completeness}). We see that the intrinsic period distribution is almost flat over $100-1000\,$d, rising gradually toward shorter periods. On the leftmost panel, we also plot the distribution of all other (i.e. non-WD+MS) binaries in red.}
    \label{fig:wdms_binaries_all}
\end{figure*}

\begin{figure*}
    \centering
    \includegraphics[width=0.9\linewidth]{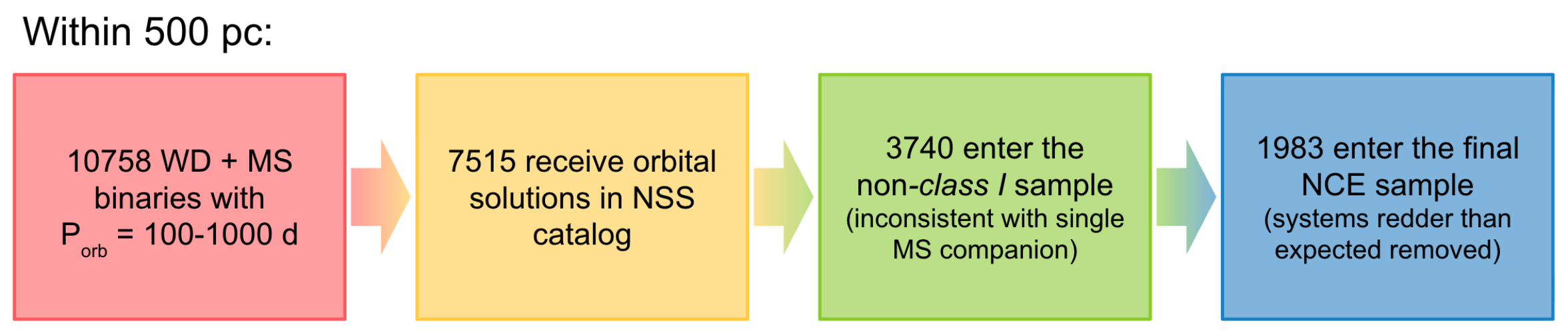}
    \caption{Total number of simulated WD+MS binaries with $P_{\rm orb} = 100-1000\,$d within 500 pc at each step in the processing pipeline.}
    \label{fig:total_numbers_flowchart}
\end{figure*}

In Figure \ref{fig:wdms_binaries_all}, we plot distributions of the orbital period and component masses of the intrinsic WD+MS binary population with $P_{\rm orb} = 100 - 1000\,$d, the majority of which are post-interaction systems. We also add plots of the resulting mock NSS catalog and the mock non-\textit{class I} and NCE samples. The shaded region around the intrinsic period distribution represents the uncertainties estimated as the ratio of the square root of the number of systems in the NCE sample (i.e. Poisson error) over the completeness fraction in each period bin (Section \ref{sssec:completeness_fracs}). In Figure \ref{fig:total_numbers_flowchart}, we provide a flowchart showing the total numbers of WD+MS binaries that make it through each step of the pipeline. 

\begin{figure}
    \centering
    \includegraphics[width=0.99\linewidth]{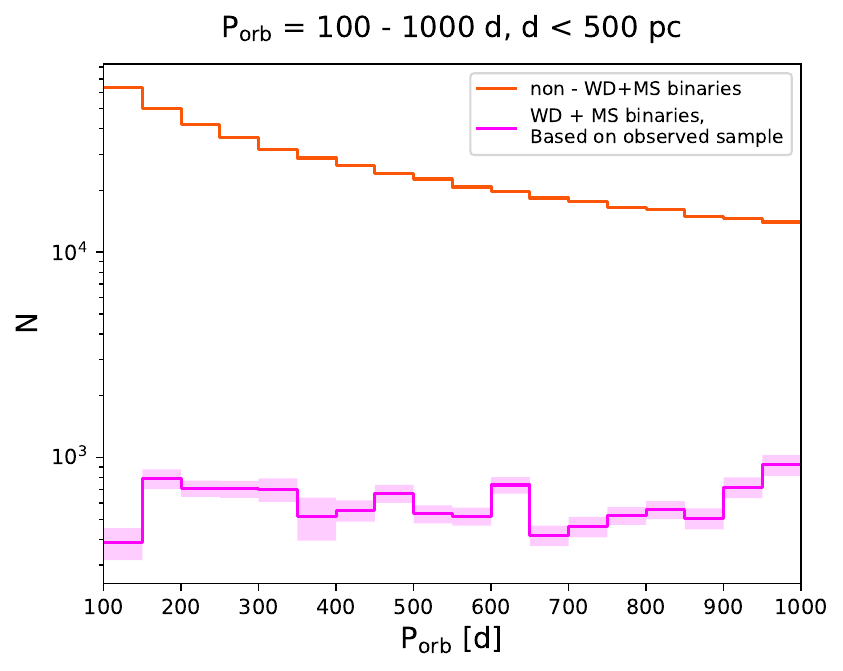}
    \caption{Comparison of the inferred intrinsic period distribution of WD+MS binaries within 500 pc (magenta) to the underlying period distribution of MS+MS binaries. At periods of 100 to 1000 d, both distributions are relatively flat, implying that the cumulative effects of interactions on the period distribution of WD+MS binaries that remain at $P_{\rm orb} > 100$ d are minor.   }
    \label{fig:wdms_binaries_all_2}
\end{figure}

The observed orbital period distribution largely reflects the completeness fraction. Once again, there is a dip at $\sim 1\,$year and a drop towards shorter periods; these features are a result of the lower sensitivity of astrometry to these orbits. The intrinsic distribution has a shallow power law slope, $\mathrm{d}N/\mathrm{d}P_{\rm orb} \propto P_{\rm orb}^{-0.13\pm0.01}$ (the error in the slope is calculated from the residuals to the fit), increasing towards shorter periods. Relative to the intrinsic population, there are fewer low-mass luminous stars in the NSS catalog. This reflects the fact that low-mass stars are faint and only yield high-quality orbits at close distances. As discussed in Section \ref{ssec:results_empirical}, cuts based on the AMRF and color excess also preferentially remove higher-mass primaries. This is because more massive stars require higher mass companions to eliminate the possibility of a luminous secondary. They also evolve more quickly, making them appear redder than expected assuming a fixed age. Few WDs at the lowest masses $\lesssim 0.5\,M_{\odot}$ enter the final sample. One reason for this is that for a given primary mass, lower mass WDs will produce smaller astrometric wobbles and are thus less likely to receive orbital solutions. Moreover, for a system with a given primary mass, those with low mass WD companions are less likely to be unambiguously identified as having non-luminous companions (i.e. most will not make it into the non-\textit{class I} sample). However, as there is a positive correlation between the masses of components in the zero-age binary population and massive primaries are disfavored by the AMRF cut, this last effect evidently does not dominate at high WD masses.

In the leftmost panel, in magenta, we plot the period distribution of the observed NCE sample divided by the completeness fraction inferred using our mock NCE sample. As expected, this closely matches the  ``intrinsic" curve, though it is slightly flatter with $\mathrm{d}N/\mathrm{d}P_{\rm orb} \propto P_{\rm orb}^{0.12 \pm 0.08}$. There is some deviation at the shortest ($\lesssim 100\,$d) and longest ($\gtrsim 900\,$d) periods which simply reflects the inconsistencies between our empirical model and observed NCE samples as seen on bottom left panel of Figure \ref{fig:nce_dist}. 

In total, in our simulated binary population, there are $95$ WDs around solar-type stars (which are taken to have masses between $0.8-1.2\,M_{\odot}$) with $P_{\rm orb} = 100 - 1000\,$d, located within 100 pc. There are $\sim 330,000$ Gaia sources within this volume  (which is almost complete; \citealt{2021A&A...649A...6G}). From this, selecting those with extinction-corrected G band absolute magnitudes between $3.0 - 6.2$ and BP-RP between $0.6 - 1.1$ (based on MIST isochrones for MS stars of masses $0.8$ and $1.2\,M_{\odot}$), we find that $\sim 8.2\%$ are solar-type stars. Given this, we calculate that $\sim 0.4\%$ of solar-type stars have WD companions at these orbital periods. This is roughly consistent with the rate of these systems inferred from WD self-lensing binaries discovered with Kepler of $(1.1\pm 0.6)\%$ \citep{Masuda2019ApJL, Yamaguchi2024MNRAS}. 

In Figure \ref{fig:wdms_binaries_all_2}, we compare the intrinsic period distributions of MS+MS binaries and WD+MS binaries within 500 pc. The WD+MS period distribution is somewhat flatter and decreases less  toward long periods, but it is still remarkably similar to the MS+MS period distribution. Within this period range, there is no hint of a period ``gap'' within which systems were cleared out by binary interactions.

\section{Conclusion} \label{sec:conclusion}


Using orbital solutions from the Gaia NSS catalog, \citet{Shahaf2024MNRAS} assembled a sample of over 3000 high-probability WD+MS binary candidates in AU-scale orbits. The formation histories of these systems are not well understood, motivating a population-level study to better understand their occurrence rates and population demographics. However, this requires an understanding of the selection function that results from both the construction of the Gaia NSS catalog and the cuts employed by \citet{Shahaf2024MNRAS} to select WD+MS binary candidates.

In this work, we use a forward modeling approach to characterize the underlying population of WD+MS binaries, extending the work of \citetalias{El-Badry2024OJAp_b}. We summarize the key findings below: 
\begin{itemize}
    \item We make modifications to the \citetalias{El-Badry2024OJAp_b} model for the evolution of binaries after binary interactions. In particular, we use MIST stellar models to impose more realistic bounds for interaction with donors on the RGB or AGB, incorporate triples into the model, and account for orbital evolution due to mass loss and binary mass transfer. 
    We tested several physically motivated models to produce the observed AU-scale WD+MS binaries (Section \ref{ssec:fshrink_models}): (1) Combining critical mass ratios from \citet{Temmink2023AA} and orbital evolution via stable MT predicted by \citet{Soberman1997AA}, (2) Using the $\alpha$-formalism for common envelope evolution to produce wide PCEBs, and (3) Using the $\gamma$-formalism, which is based on angular momentum conservation. 
    \item We also constructed an empirical model tuned to reproduce the observed sample. We set a fixed value for the critical mass ratio of $M_{\rm accretor}/M_{\rm donor} = 0.38$ for systems with AGB donors. Above this, we implemented the \citet{Soberman1997AA} relation assuming fully non-conservative mass transfer. All other interacting systems are assumed to shrink to small orbits (Section \ref{sssec:empirical_model}).
    \item The model for stable MT based on \citet{Temmink2023AA} overproduces the number of au-scale WD+MS binaries by a factor of $\sim 3$. It produces a period distribution in reasonable agreement with observations, but it is unable to reproduce the deficit of high-mass WDs above $\sim 0.8\,M_{\odot}$ or the narrow distribution of companion MS star masses. We conclude that binaries in nature are on average less stable than predicted by the model (Section \ref{ssec:results_T23}).
    \item Using the $\alpha$-formalism prescription to evolve orbits, we require very large $\alpha\lambda \sim 10^4$ for $\sim 10\%$ of systems that interact with an AGB donor to form sufficiently wide WD+MS binaries. However, the final sample resulting from this model overpredicts the proportion of short period ($P_{\rm orb} \lesssim 1\,$yr) systems and does not reproduce the deficit of high-mass WDs (Figure \ref{fig:porb_mwd}, Section \ref{ssec:results_alpha_lambda}). Moreover, it is unclear why such a high $\alpha\lambda$ value would apply to only $\sim10\%$ of AGB donors. We conclude that the formation of wide WD+MS binaries cannot be easily understood through this formalism. We find that the resulting sample from the model based on the $\gamma$-formalism with $\gamma = 2.0$ suffers similar issues (Section \ref{ssec:results_gamma}).
    \item The simulated sample of astrometric WD+MS binaries obtained from our empirical model broadly reproduces the observed period and component mass distributions (Figure \ref{fig:nce_dist}). $14\%$ of systems in our mock NSS catalog are WD+MS binaries, out of which $\sim15\%$ enter the final NCE sample. 
    \item In our empirical model, the binaries that do not remain wide are those which have small accretor-to-donor mass ratios. A steep dependence of the amount of orbital shrinkage on the initial mass ratio is required to produce the strongly peaked WD mass distribution seen in the observed sample (Section \ref{sec:results_modified}). 
    \item While a significant number of hierarchical triples may enter the non-\textit{class I} sample, the cut on color excess removes the vast majority of them. Therefore, we expect the final WD+MS binary sample selected by \citet{Shahaf2024MNRAS} to be pure. However, our modeling predicts that about half of WD+MS binaries that receive orbital solutions are excluded by the color excess cut. We also find that this cut likely causes the observed NCE sample to be biased against metal-poor systems. A less stringent selection, perhaps one with a dependence on metallicity, may be implemented so that fewer true WD+MS binaries are removed.  
    \item The post-MT eccentricities of AU-scale WD+MS binaries are typically between 0 and 0.2. Completeness does not significantly affect this distribution. (Section \ref{ssec:eccentricity_evol}, Figure \ref{fig:nce_dist}) 
    \item The completeness of the Gaia WD+MS binary sample as a function of orbital period shows a characteristic dip at 1 yr due to the degeneracy between parallactic and orbital motion and decreases towards the outskirts at $\sim 100$ and $1000\,$d. Taking this into account, we find that the intrinsic orbital period distribution of WD+MS binaries between this period range is close to flat, where $\mathrm{d}N/\mathrm{d}P_{\rm orb} \propto P_{\rm orb}^{0.12 \pm 0.08}$.  
    \item The completeness varies non-trivially with the mass of the primary and WD (Figure \ref{fig:completeness}). This is because while more massive primaries are on average more luminous and thus preferentially receive orbital solutions, they are disfavored in ruling out luminous secondaries. Meanwhile, at a given primary mass, there is a bias towards more massive WDs but this is offset by the intrinsic correlation that exists between the component masses. 
    \item Our model implies that around $0.4\%$ of all solar-type stars have WD companions in with $P_{\rm orb} \sim 100-1000\,$d, consistent within about a factor of two with constraints from the observed number of WD self-lensing binaries. 
\end{itemize}

\section{Acknowledgements}

We thank the referee for providing detailed feedback which greatly improved the manuscript. 

This research was supported by NSF grant AST-2307232. NY acknowledges support from the Ezoe Memorial Recruit Foundation scholarship.

This work has made use of data from the European Space Agency (ESA) mission {\it Gaia} (\url{https://www.cosmos.esa.int/gaia}), processed by the {\it Gaia} Data Processing and Analysis Consortium (DPAC,
\url{https://www.cosmos.esa.int/web/gaia/dpac/consortium}). Funding for the DPAC has been provided by national institutions, in particular the institutions participating in the {\it Gaia} Multilateral Agreement.

\vspace{5mm}
\facilities{Gaia}

\software{gaiamock \citep{El-Badry2024OJAp_b}, stam \citep{Hallakoun2021MNRAS}}

\appendix

\section{Color excess} \label{appendix:CE_ecc_metallicity}

\subsection{Relation between triple fraction and metallicity} \label{ssec:appendix_triplefrac_met}

\begin{figure}
    \centering
    \includegraphics[width=0.45\linewidth]{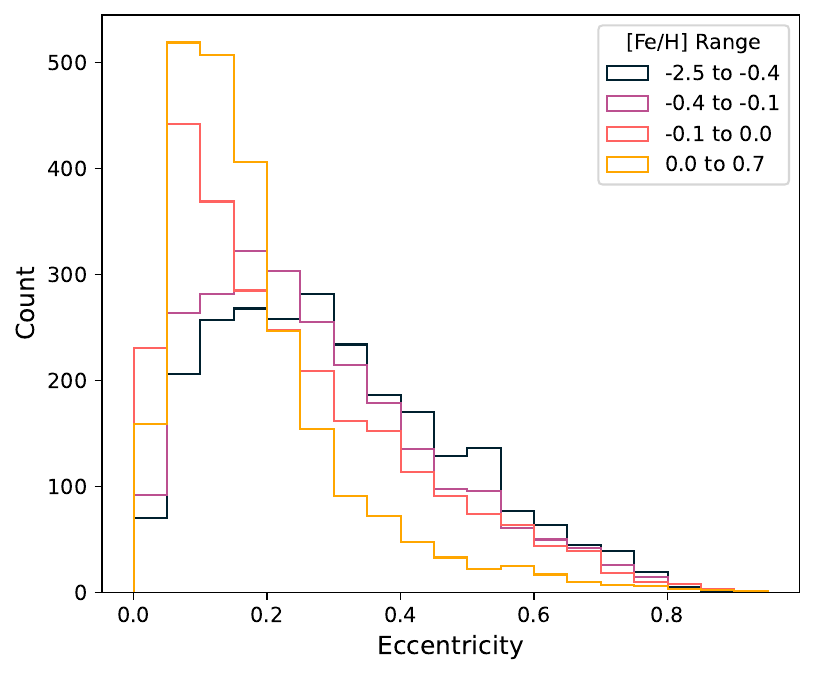}
    \caption{Eccentricity distribution of the \citet{Shahaf2024MNRAS} non-\textit{class I} sample grouped by [Fe/H]. The ranges are chosen such that there is roughly the same total number in each group. For [M/H] $>$ -0.1, there is an excess of low-eccentricity systems, likely due to WD+MS binaries which have undergone some circularization. For lower metallicities, the eccentricities are higher, possibly reflecting a higher fraction of triples.}
    \label{fig:ecc_feh_dist}
\end{figure}

In Section \ref{ssec:color_excess_cut}, we noted the presence of a correlation between color excess and eccentricity in the observed non-\textit{class I} sample (Figure \ref{fig:real_CE_v_params}). Since we generally expect triples to have higher eccentricities than WD+MS binaries, this may be attributed to a higher triple fraction at large color excess. Figure \ref{fig:real_CE_v_params} also shows a strong anti-correlation between color excess and metallicity. This could reflect (a) deficiencies in the color excess metric that are more important for old and metal-poor stars, (b) an increased triple fraction at low metallicity, or (c) a bias in the measured metallicities for triples. To discriminate between these possibilities, we investigate the eccentricity distributions of sources in different color excess ranges.

In Figure \ref{fig:ecc_feh_dist}, we plot the eccentricity distribution of sources in the non-\textit{class I} sample binned in several different metallicity bins. The eccentricity distributions of the more metal-poor samples are broader than those of metal-rich samples, with more eccentric systems making up a larger portion of the population. This suggests that the trend in color excess with metallicity is not simply a result of biased color excess measurements at low metallicity, as in this case one would expect a low-eccentricity population to also be present at low metallicities. A simpler explanation is that the presence of light from unmodeled secondaries and tertiaries may bias the metallicity measurements from \citet{Zhang2023MNRAS} toward lower metallicities. It is also possible that compact triples are simply more common at low metallicity. We defer further exploration of the metallicity -- color excess anti-correlation to future work.

\subsection{Empirical color excess correction to models} \label{ssec:appendix_CE_correction}

\begin{figure}
    \centering
    \includegraphics[width=0.99\linewidth]{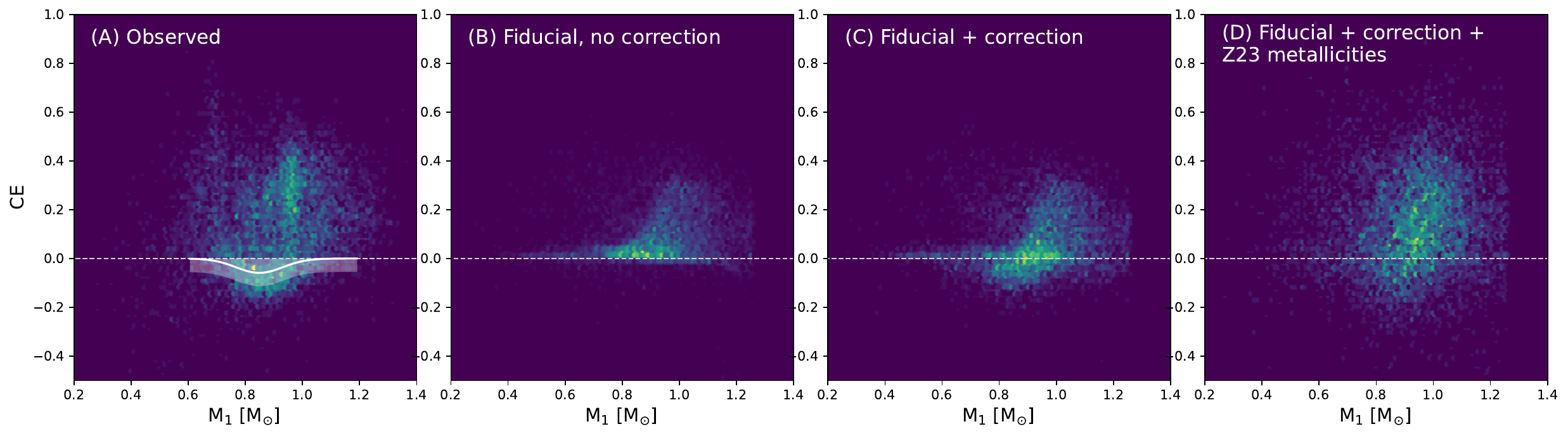}
    \caption{2D binned plots of color excess against the luminous primary mass of the non-\textit{class I} systems. From left to right, we have the (\textit{A}) observed  \citet{Shahaf2024MNRAS} sample with the fitted Gaussian curve used for the empirical correction (Section \ref{ssec:appendix_CE_correction}) (\textit{B}): our empirical model with no corrections made in the calculation of color excess, (\textit{C}): same as (B) but with the empirical correction on color excess applied, and (\textit{D}): same as (C) but with metallicities drawn from the observed \citet{Zhang2023MNRAS} distribution (Section \ref{ssec:appendix_CE_met_spread}).}
    \label{fig:color_excess_correction}
\end{figure}

We apply an empirical correction to the color excess of our modeled systems to reproduce the observed trend between color excess and primary mass. In theory, true WD+MS binaries are expected to have color excess values close to zero, with more scatter toward positive values due to systems in which the MS star is older than 2 Gyr and has evolved significantly. Therefore, the observed offset towards negative color excess is indicative of a mismatch between the observed colors and those predicted by PARSEC isochrones. We fit the observed color excess values for sources with $M_1 = 0.7 - 1.2\,M_{\odot}$ and color excess $<$ 0.1 (left panel of Figure \ref{fig:color_excess_correction}) with a Gaussian. To reproduce the observed scatter, we also add Gaussian noise with a standard deviation of 0.05 to the simulated color excess values. The fit to the observed non-\textit{class I} sample is shown on panel A of Figure \ref{fig:color_excess_correction}. Panels B and C show the systems in our empirical model before and after the correction, respectively. 

\subsection{Discrepancy in the color excess distribution} \label{ssec:appendix_CE_met_spread}

\begin{figure}
    \centering
    \includegraphics[width=0.85\linewidth]{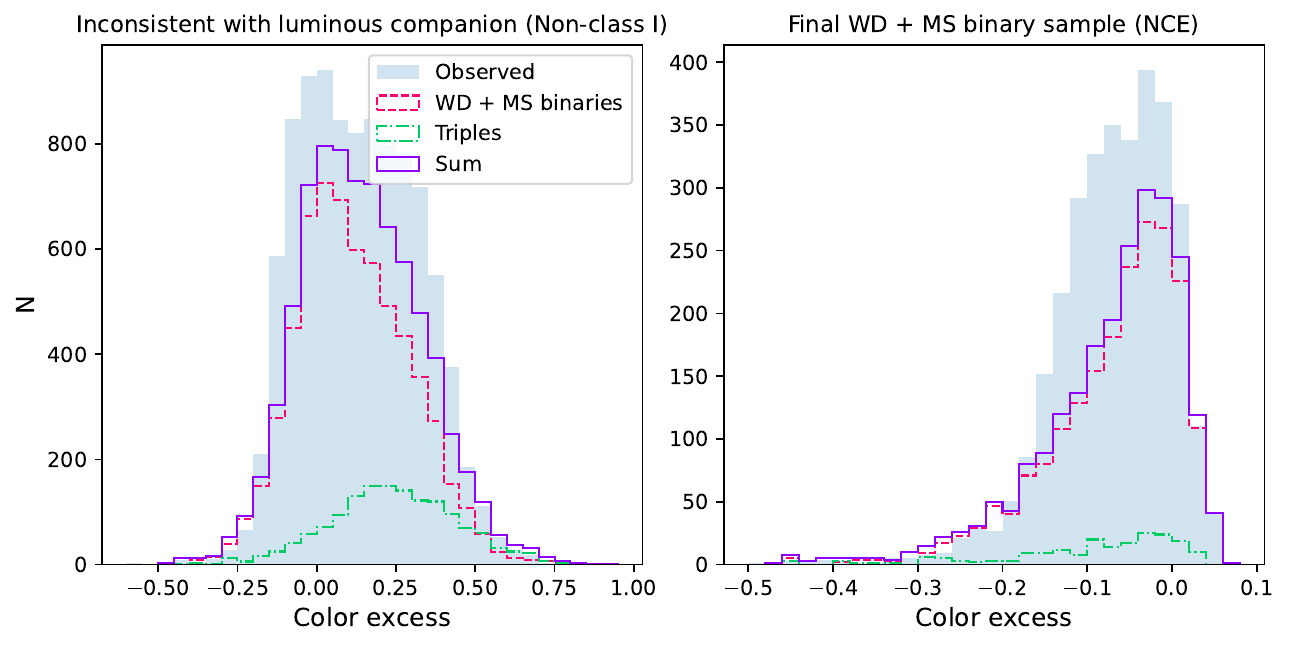}
    \caption{Analogous to Figure \ref{fig:color_excess_dist}, but with metallicities used to calculate $(B-I)_{\rm expected}$ for the modeled systems drawn from the \citet{Zhang2023MNRAS} metallicity distribution of the observed non-\textit{class I} sample. As expected, the discrepancy introduced between these mock ``measured" metallicities  and true values used to calculate $(B-I)_{\rm observed}$ leads to broader distributions (c.f. Figure \ref{fig:color_excess_dist}) which better match the observed sample.}
    \label{fig:color_excess_dist_Z23_fehs}
\end{figure}

As seen in Figure \ref{fig:color_excess_dist}, the distribution of color excess is significantly more peaked around zero for our model compared to the observed sample. A potential cause is that the measured metallicities either (a) do not reflect the true values or (b) are drawn from a broader distribution than assumed in our Galaxia simulations. Here, we test this possibility by randomly drawing metallicities and their uncertainties from the observed non-\textit{class I} sample in calculating $(B-I)_{\rm expected}$. Similarly to our empirical model, we once again select 20\% of the systems to have solar metallicity with an error of 0.25 dex. In Figure \ref{fig:color_excess_dist_Z23_fehs}, we show the color excess distributions of the non-\textit{class I} and NCE samples obtained after this change has been made. The mismatch introduced between the true and ``measured" metallicities significantly broadens the color excess distribution compared to that in Figure \ref{fig:color_excess_dist} so that it is in better agreement with the observed sample. The broadening is also seen in panel D of Figure \ref{fig:color_excess_correction}, which shows a 2D histogram of color excess and primary mass. We conclude that biases in the inferred metallicities for triples could plausibly explain both the broadness of the observed color excess distribution and the observed trend in color excess with metallicity. This possibility is also appealing because the metallicity distribution of the observed non-class I sample is significantly broader, and centered on more metal-poor values, than the metallicity distribution of all stars in the solar neighborhood \citep[e.g.][]{Hayden2015ApJ}. However, we emphasize that this has minimal effect on the distributions of key stellar and orbital parameters in the final WD+MS binary sample plotted in Figure \ref{fig:nce_dist}, because the triples with potentially biased metallicities are ultimately excluded from the final WD+MS binary sample.

\subsection{WD IFMR} \label{ssec:appendix_mod_wd_ifmr}

\begin{figure*}
    \centering
    \includegraphics[width=0.5\linewidth]{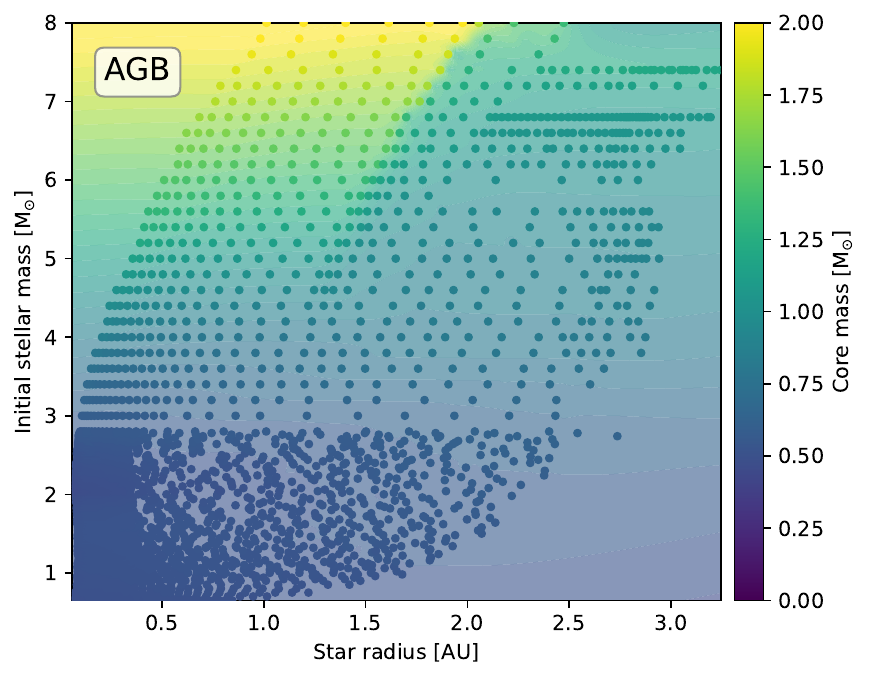}
    \includegraphics[width=0.45\linewidth]{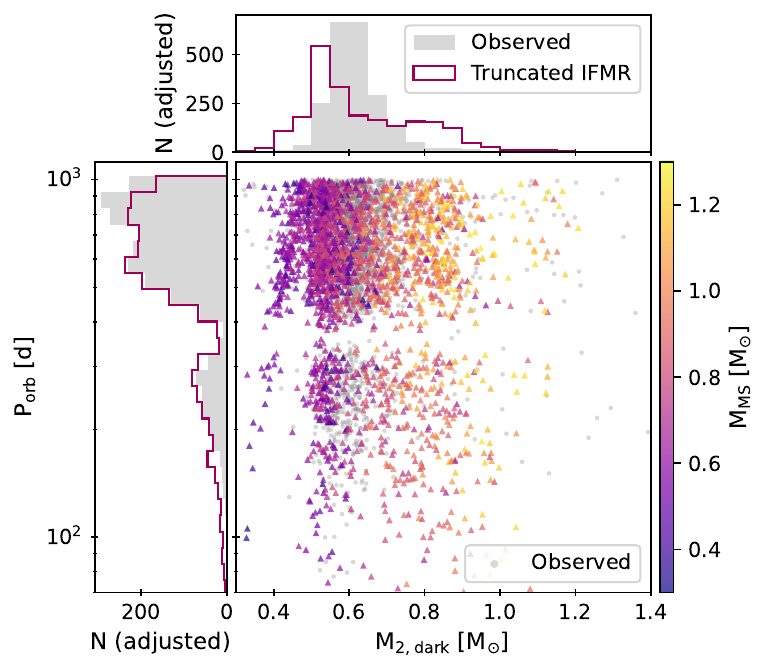}
    \caption{\textit{Left}: Core mass as a function of initial mass and radius on the AGB. The markers are the data points from MIST tracks used in the interpolation. \textit{Right}: The resulting astrometric sample of WD+MS binary candidates on the $P_{\rm orb}-M_{\rm 2, dark}$ from using the IFMR derived from interpolating the core masses in the left panel. A detailed description of the plot can be found in the text under Figure \ref{fig:porb_mwd}.}
    \label{fig:trunc_IFMR}
\end{figure*}

We also test the effect of using a different WD IFMR which takes into account the interruption of the progenitor's evolution due to the onset of binary interaction. This may be particularly important if the system undergoes CEE and the envelope is quickly ejected. On average, this should lead to lower WD masses compared to those predicted from single star evolution and possibly ameliorate the discrepancy between the observed and simulated WD mass distributions (Section \ref{sec:e24_model}). We estimate WD mass as the \texttt{he\_core\_mass} from the MIST evolutionary tracks at a given giant radius and initial mass (Left panel of Figure \ref{fig:trunc_IFMR}; note that these cores are primarily carbon/oxygen by mass -- the MESA label only indicates that the outer helium layer is included). We used the same method for the interpolation to predict the primary mass at the onset of MT (Section \ref{sssec:T23_model}). We see that the core mass can exceed $1.4\,M_{\odot}$ at the final stages of the AGB phase for progenitor masses $\sim 6-8\,M_{\odot}$. In this case, we return to using the \citet{Weidemann2000AA} IFMR. This is reasonable as the core is defined in terms of the helium mass fraction and not all of this material ends up burning into carbon and oxygen. A very small fraction of systems are in this regime so the end results are insensitive to this choice. 

For stars that interact on the AGB, this IFMR leads to WD masses that are $\sim 10-20\%$ lower than those predicted by the \citet{Weidemann2000AA} relation for progenitor masses between $\sim 1.5$ and $4.5\,M_{\odot}$. This corresponds to WD masses between $\sim 0.55-0.85\,M_{\odot}$. Outside this range, our IFMR can lead to higher WD masses. For comparison, \citet{Shahaf2025ApJ} used a different method on a spectroscopic sample of AU-scale WD + red giant binaries in open clusters and found that these WDs were $\sim 20\%$ less massive than isolated WDs, which they attributed to binary interaction (also see \citealt{Ironi2025ApJ}). 

The final WD+MS binary sample resulting from using this modified IFMR with the stable MT model (Section \ref{sssec:T23_model}) is plotted on the $P_{\rm orb}-M_{\rm 2, dark}$ space in the right panel of Figure \ref{fig:trunc_IFMR}. All other assumptions are the same as in our empirical model (Section \ref{sssec:empirical_model}). We see that while the orbital period is not dramatically affected, the WD mass distribution now peaks at $\sim 0.55\,M_{\odot}$, below that of the observed sample. Furthermore, there remains too many WDs above $\sim 0.8\,M_{\odot}$. 

\bibliographystyle{aasjournal}



\end{document}